# Theory of Photo-injection of Hot Plasmonic Carriers from Metal Nanostructures into Semiconductors and Surface Molecules

Alexander O. Govorov[1]*, Hui Zhang[1], Yurii K. Gun'ko[2]

[1]*Department of Physics and Astronomy, Ohio University, Athens, Ohio 45701*

[2]*School of Chemistry, University of Dublin, Trinity College, Dublin 2, Ireland*

## ABSTRACT

We investigate theoretically the effects of generation and injection of plasmonic carriers from an optically-excited metal nanocrystal to a semiconductor contact or to surface molecules. The energy distributions of optically-excited hot carriers are dramatically different in metal nanocrystals with large and small sizes. In large nanocrystals, the majority of hot carriers have very small excitation energies and the excited-carrier distribution resembles the case of a plasmon wave in bulk. For nanocrystal sizes smaller than 20nm, the carrier distribution extends to larger energies and occupies the whole region $E_F < \varepsilon < E_F + \hbar\omega$. The physical reason for the above behaviors is non-conservation of momentum in a nanocrystal. Because of the above properties, nanocrystals of small sizes are most suitable for designing of opto-electronic and photosynthetic devices based on injection of plasmonic electrons and holes. For gold nanocrystals, the optimal sizes for efficient generation of hot carriers with over-barrier energies are in the range of 10-20nm. Another important factor is the polarization of the exciting light. For efficient excitation of carriers with high energies, the electric-field polarization vector should be perpendicular to a prism-like nanoantenna (slab or platelet). We also show the relation between our theory for injection from plasmonic nanocrystals and the Fowler theory of injection from a bulk metal. Along with a prism geometry (or platelet geometry), we consider cubes. The results can be applied to design both purely solid-state opto-electronic devices and systems for photo-catalysis and solar-energy conversion.

*Corresponding Author: Govorov@ohiou.edu

**Introduction.**

Injection of hot plasmonic carriers from metal nanostructures and nanocrystals is a very interesting and active direction of research. It has direct relevance to solid-state devices[1], photo-chemical systems, sensors, and solar cells[2]. Plasmonics is particularly attractive for hot carrier injection. Metal nanocrystals have large absorption cross sections and can enhance and trap light[3,4]. Metal has a large density of electrons and these electrons can be used for surface chemical reactions[5,6,7,8,9,10,11,12,13,14,15] (when nanocrystals are in a solution) or for photo-injection of carriers in solid-state nano-devices[16,17,18,19,20,21]. In the standard picture of the photo-injection in a semiconductor-metal Schottky-barrier device, photo-excited electrons in a metal have an isotropic distribution in the momentum space; then, only the electrons within the momentum cone can propagate into the semiconductor contact. This is a central assumption of the Fowler theory[22]. However, very recent experiments on injection from plasmonic nano-antennas suggested that hot carrier distributions in plasmonic nanoscale systems may have directionality in the momentum space[23]. Several theoretical papers developed further the Fowler theory considering: (1) reflection of electron waves in slabs[24,25] and (2) plasmonic effects[26,27,28]. The quantum formulas for surface photo-electric effect[29,30] were used to calculate the photo-injection efficiencies in plasmonic nanostructures in Refs.[31,32]. A recent paper[33] introduced the effect of effective temperature in the non-equilibrium steady state of optically-driven metal nanoparticle.

Here we employ a quantum theory based on the equation of motion of density matrix to calculate the distribution of optically-excited electrons and holes in a metal nanocrystal. Using this distribution, we evaluate injection efficiencies of plasmonic nanostructures. For this, we assume two experimental schemes: Injection into a semiconductor contact and transfer of electrons to surface molecules. Importantly, efficient generation of hot electrons with large energies appear only in small nanocrystals with dimensions below 20 nm. A small size of a nanocrystal is needed to break down the momentum conservation in the process of interaction between phonons and confined electrons. In large nanocrystals, the majority of hot electrons has very small excitation energies (i.e. the energies of excited electrons are close to the Fermi level) and the number of highly-excited carriers is small and can originate from defects. The direction of light polarization is also very important in small nanocrystals. In a prism-like antenna (platelet), the polarization of exciting field should be along the shortest axis; in this case, the number of highly-excited electrons is expected to be large. Other shapes suitable for generation of hot carriers are cubes and nanowires. In the end of the paper, we discuss the Fowler theory applicable for a bulk metal and how it relates to our results derived for relatively small nanocrystals. In contrast to the previous calculations of photoelectric effect in plasmonic nanocrystals[24,26,31,32,33] we developed a quantum microscopic picture of non-equilibrium carrier population in a localized plasmon wave using the self-consistent (Hartree-like) approach and show that the energy distribution of hot plasmonic carriers strongly depends on a size of a nanocrystal. Our paper is also relevant to the sub-field of plasmonics studying quantum[34,35,36] and quasi-classical effects[37] in metal nanocrystals.

## 1. Formalism and models.

Figure 1 shows a schematics of the systems of interest. Incident photons excite a Fermi gas of electrons in a metal nanocrystal (**NC**) and some of the electrons become transferred to a semiconductor contact (Fig. 1a-c) or to molecules (Fig. 1d). To describe the electronic state of NC, we employ here a method of the equation of motion of density matrix $\hat{\rho}$. The equation of motion has the form[38,39]

$$\hbar \frac{\partial \rho_{mn}}{\partial t} = i \langle m | [\hat{\rho}, \hat{H}_0 + \hat{w}_{tun} + \hat{V}_{opt}] | n \rangle - \Gamma_{mn} (\hat{\rho}_{mn} - \hat{\rho}_{mn}^{(0)}), \qquad (1)$$

where $\rho_{mn}^{(0)} = f_n^0(\varepsilon_n) \cdot \delta_{nm}$ is the equilibrium matrix involving the Fermi distribution function $f_n^0 = f_F(\varepsilon_n)$, $\Gamma_{nm} = \hbar / \tau_{nm}$ are the relaxation rates. The Hamiltonian contains three terms described in the next lines. The operator $\hat{H}_0$ is the single-particle Hamiltonian of the isolated components. These components include a confined nanocrystal (NC) and a semi-infinite semiconductor contact. Or, the latter can be also a molecule in a solution. The operator $\hat{w}_{tun}$ describes the tunneling between a NC and a contact (or a surface molecule for a NC in a solution). The third term is the light-matter interaction in the electron plasma:

$$\hat{V}_{opt} = \sum_i V_\omega(r_i) e^{-i\omega t} + V_\omega^*(r_i) e^{i\omega t},$$
$$V_\omega(r) = e\varphi_\omega(r) = e[\varphi_{0,\omega}(r) + \varphi_{ind,\omega}(r)],$$

where the sum is taken over all electrons in the system and $\varphi_{0,\omega}$ is a potential induced by an incident monochromatic wave and $\varphi_{ind,\omega}$ is the induced Coulomb potential in the plasma. This approach is equivalent to the so-called random-phase approximation method in the theory of many-body systems[38].

In the linear regime discussed in the paper, the off-diagonal components of the density matrix are linear and are given by:

$$\rho_{nm}(t) = \rho_{nm,a} e^{-i\omega t} + \rho_{nm,b} e^{i\omega t},$$
$$(\hbar\omega - \varepsilon_n + \varepsilon_m + i\Gamma_{nm})\rho_{nm,a} + \sum_{\alpha}(\rho_{n\alpha}w_{\alpha m} - w_{n\alpha}\rho_{\alpha m}) = -e\varphi_{nm,a}(f_n^0 - f_m^0), \quad (2)$$
$$(-\hbar\omega - \varepsilon_n + \varepsilon_m + i\Gamma_{nm})\rho_{nm,b} + \sum_{\alpha}(\rho_{n\alpha}w_{\alpha m} - w_{n\alpha}\rho_{\alpha m}) = -e\varphi_{nm,b}(f_n^0 - f_m^0),$$

where the indices $n$ and $m$ relate to the metal NC, $\varepsilon_n$ are single-electron energies, $f_n^0 = f_F(\varepsilon_n)$, and $\varphi_{nm,a} = \langle n|\varphi_\omega(r)|m\rangle$ and $\varphi_{nm,b} = \langle n|\varphi_\omega^*(r)|m\rangle$. The index $\alpha$ belongs to the semiconductor contact (or to a molecule) and $w_{n\alpha} = \langle n|\hat{w}_{tun}|\alpha\rangle$ are the tunnel matrix elements. The diagonal elements of the density matrix are non-linear:

$$\delta\rho_{nn} = \rho_{nn} - \rho_{nn}^{(0)} = -\frac{2e}{\Gamma_{nn}} \text{Im}\left(\sum_{n'}\rho_{nn',a}\varphi_{n'n,b} + \rho_{nn',b}\varphi_{n'n,a} + \sum_{\alpha}\rho_{n\alpha}w_{\alpha n}\right), \quad (3)$$

where $\rho_{nn}^{(0)} = f_n^0 = f_F(\varepsilon_n)$. Since the tunneling makes the problem very complex in many cases, it is useful to solve Eq. 2 first for an isolated system. In the following steps, we include the interactions with the leads or molecules using various approximations (Supplementary information). Omitting the tunneling terms in Eqs. 2 and 3, the off-diagonal and diagonal matrix elements of the operator $\hat{\rho}$ can be found in the following form:

$$\rho_{nn',a} = e\varphi_{nn',a}\frac{(f_{n'}-f_n)}{(\hbar\omega - \varepsilon_n + \varepsilon_{n'} + i\Gamma_{nn'})}, \quad \rho_{nn',b} = e\varphi_{nn',b}\frac{(f_{n'}-f_n)}{(-\hbar\omega - \varepsilon_n + \varepsilon_{n'} + i\Gamma_{nn'})},$$

$$\delta\rho_{nn} = \rho_{nn} - \rho_{nn}^{(0)} = \frac{2e^2}{\Gamma_{nn}}\sum_{n'}(f_{n'}^0 - f_n^0)\left[|\varphi_{nn',a}|^2\frac{\Gamma_{nn'}}{(\hbar\omega-\varepsilon_n+\varepsilon_{n'})^2+\Gamma_{nn'}^2} + |\varphi_{nn',b}|^2\frac{\Gamma_{nn'}}{(\hbar\omega+\varepsilon_n-\varepsilon_{n'})^2+\Gamma_{nn'}^2}\right].$$

(4)

The numbers of excited electrons and holes can be now calculated from these equations for both unbounded and bounded systems. For simplicity we will take $\Gamma_{nn'} = \Gamma$ in the following calculations.

## 2. Unbounded metal systems. Wave function of a plasmon and a distribution of hot electrons in the plasmon wave.

It is instructive to start from plasmon excitations in unbounded 3D systems. In the 3D case, the solution of the equation of motion of the density matrix within the self-consistent field approximation (random-phase approximation) is exact and simple[38]. The wave function of a plasmon wave can be revealed from the elements of the density matrix:

$$\rho_{mn}(t) = \langle \Psi(t)|\hat{c}_n^+\hat{c}_m|\Psi(t)\rangle,$$

The presence of nonzero element $\rho_{mn}(t)$ means that the Fermi gas has an excitation: An electron-hole pair shown in Figure 1b. In this excitation, an electron occupies in the exciton state $|m\rangle$ above the Fermi level and a hole appears in the state $|n\rangle$ below the Fermi energy.

**2.1 Distributions of hot electrons and holes in the plasmon wave. Free electrons.**

It is interesting to look at the hot-electron distribution in a bulk plasmon wave. Using the random-phase approximation, one can solve this problem[38]. The details of the solution can be found in Supporting Information. The key parameter describing the response of Fermi gas is the dielectric function:

$$\varepsilon_{3D}(\omega, q) = 1 - \frac{4\pi e^2}{Vq^2} \sum_k \frac{f^0_{\mathbf{k-q}} - f^0_{\mathbf{k}}}{\hbar\omega - \varepsilon_{\mathbf{k}} + \varepsilon_{\mathbf{k-q}} + i\Gamma}.$$

In this function, we see the factor $f^0_{\mathbf{k-q}} - f^0_{\mathbf{k}}$ that shows the character of electronic transition sin the Fermi gas. Transitions occur in the vicinity of the Fermi level and lead to creation of electron-hole pairs (Figure 2b). In most cases for bulk plasmons, $\hbar q \cdot v_F \ll \omega_p$ and $q \ll k_F$. Then, for these conditions, the hot-electron distribution takes the following form:

$$\delta n_e(\varepsilon) = A \cdot \frac{\varepsilon_F - \varepsilon + \hbar q v_F}{\hbar q v_F}, \quad \varepsilon_F < \varepsilon < \varepsilon_F + \hbar q v_F,$$

$$A = V \sqrt{\frac{2m\varepsilon_F}{\hbar^2}} \frac{2m}{\pi^2 \hbar^2} \frac{e^2 |\tilde{\varphi}_{\omega,q}|^2}{(\hbar \omega_p)^2}.$$

(5)

From this equation, we again see that hot electrons are distributed in a relatively narrow region of energies $\varepsilon_F < \varepsilon < \varepsilon_F + \hbar k v_F$.

In the same manner, we can derive the distribution of hot holes shown as a region with red dots in Figure 2b. In the case, we look at the energies below the Fermi level. In the k-space, the hole distribution is $f_\mathbf{k}^0 - \rho_{\mathbf{kk}}$ assuming $\varepsilon < \varepsilon_F$. We again integrate over angles and obtain the distribution of hot holes over energies:

$$\delta n_h(\varepsilon) = \frac{\delta p_\varepsilon}{\delta \varepsilon} = A \frac{\varepsilon - \varepsilon_F + \hbar q v_F}{\hbar q v_F}, \quad \varepsilon_F - \hbar q v_F < \varepsilon < \varepsilon_F.$$

Again, holes in the plasmonic wave are distributed near the Fermi surface. A characteristic energy of hot carrier in a plasmonic wave is $\sim \hbar q v_F$.

Figures 2c) and d) show the calculated hot-carrier distributions for two plasmonic wavelengths in a monochromatic plasmon wave with a single wave vector $q_p = \frac{2\pi}{\lambda_p}$, where $\lambda_p$ is a wavelength of a plasmon. Since gold is a monovalent metal, the Fermi energy and velocity can be calculated from the free-electron theory and the mass density: $E_F = 5.5 eV$ and $v_F = 1.39 \cdot 10^8 cm/s$. Such plasmonic wave can be launched in a planar plasmonic waveguide shown in Figure 2a and described in some details in Supporting Information.

Considering a monochromatic plasmon wave with a wave vector $q_p$ and an energy $\hbar\omega_p$, we can also estimate an average number of excited carries in a single plasmon. As expected, a plasmon wave is made of a large number of electron-hole pairs:

$$\frac{\delta N_{electrons}}{\delta N_{plasmon}} = const \cdot \frac{\hbar\omega_p}{\hbar q_p v_F} \gg 1,$$

where $\delta N_{electrons}$ is an average number of excited electrons and $\delta N_{plasmons}$ is a number of excited plasmons, and $const \sim 1$. The above ration is an average number of excited electrons per one plasmon quantum and this ratio is typically a large number for bulk and surface plasmons. Details of the derivation are given in Supporting Information.

**2.2 Real metals. Intra-band and inter-band transitions in the plasmon wave.**

Figure 2e also illustrates qualitatively the distribution of electrons and holes due to the inter-band transitions in a plasmonic wave. Such distribution should be calculated numerically using the band structure of a real metal. However, it is possible to understand qualitatively the role of inter-band transitions using the known band structure. The band diagram of gold from Ref.[40] is reproduced in Figure 2f. The intra-band transitions depend on a wavevector of a plasmon and appear close to the Fermi level since a plasmon wavevector is typically small (red arrows in Figure 2f). Inter-band excitations come from transitions between different bands (blue arrows in Figure 2f) and can be calculated in the limit $q_p \to 0$.

**2.3 A plasmon wave in an Au slab.**

We now consider one illustration - a simple model case of a plasmonic slab. Surface plasmon polaritons (SPPs) can be launched in a slab of metal using optical excitation and a grating. Figure 2a shows the geometry of the system. The dispersion of the SPPs[41,42] can be calculated from the classical electrodynamics using the local dielectric function. It is striking that the hot-carrier energy are so small compared to the typical energies of surface plasmons excited in planer metal waveguides[41,42]. Simultaneously, the hot-carrier energies are much smaller than the typical Schottky barriers in the metal-semiconductor junctions[1]. Therefore, ideally, we need a short wavelength plasmon to create substantial over-barrier current. This can be achieved in nanostructures with small sizes and using directional excitation. In real devices with relatively wide waveguides, electric currents generated by a plasmon wave can be nevertheless detected[43]. This may appear, for example, because of defects and interfaces which may break conservation of momentum of electron. The corresponding currents are usefully modeled using the Fowler theory[22]. Below, in the Section 6, we will return to this discussion.

We now consider a waveguide with grating. Plasmons can be launched in such waveguide using light with the normal incidence (see Figure 2a). It is striking that the characteristic energy of hot carriers excited due to the intra-band transitions is so small (Figures 2b and c). As it was pointed out above that the characteristic energy of hot carriers is $\sim \hbar q_{SSP} v_F$. For excitation with realistic grating periods of $d = \lambda_{SPP} = 790 nm$ and $224 nm$, we obtain small energies of hot electrons, *~7* and *26 meV*. These energies are much smaller than typical energies for the Schottky barriers in Au-Si and Au-TiO$_2$ devices which are ~ 0.5eV[1,16] and 1.2eV. Therefore, the currents observed in the experiments with the Schottky junction may be induced by defects or interfaces which destroy conservation of momentum and induce high-energy excitations in an electron plasma (see the discussion in Section 6). However, in an ideal crystal, a plasmon with a short wavelength may be able to induce significant hot carrier currents. For example, a plasmon with a wavelength $\lambda_{SPP} = 10 nm$ can create intra-band hot carries with the energy ~ $\hbar q_p v_F$ ~ $0.57 eV$. Such plasmon mode can induce over-barrier currents in a Schottky junction. This simple conclusion is important and suggests that optically-excited small nanocrystals with sizes *L~10-20nm* may have a significant number of plasmonic carriers with over-barrier energies and, therefore, such nanocrystals can generate hot plasmonic injection into an electrode.

Inter-band hot carriers are also interesting and may be used for photo-currents; especially the hot inter-band holes look promising. The threshold of *active* inter-band absorption (at ~ 2.4eV) occurs in the vicinity of the L-point[40] (Figure 2f). At the threshold $\hbar\omega \sim 2.4 eV$, hot electrons are created in the vicinity of the Fermi level $\varepsilon_{e,hot} \sim \varepsilon_F$, but the hot holes are created obviously at $\varepsilon_{h,hot} \sim \varepsilon_F - \hbar\omega$ with $\hbar\omega \sim 2.4 eV$. With increasing the excitation frequency, the hot carrier energies, $\varepsilon_{e,hot}$ and $\varepsilon_{h,hot}$, are expected to increase. A semi-quantitative estimation for these energies may be derived from the following arguments. The occupied d-band is relatively flat (Fig. 2f) and, therefore, the hot hole energy in the first approximation can be fixed at $\varepsilon_{h,hot} \sim E_F - 2.4 eV$, whereas the hot-electron energy is an increasing function of $\omega$: $\varepsilon_{e,hot} = E_F - 2.4 eV + \hbar\omega$ for $\hbar\omega > 2.4 eV$. Figure 2e illustrates this situation. From the above numbers, we see that hot holes can have large energies $E_F - \varepsilon_{h,hot} \geq 2.4 eV$, i.e. the inter-band holes are generated relatively far from the Fermi energy. These energies are sufficient for injection of holes into a *p-doped* semiconductor contact and, therefore, photo-generated hot holes in the d-band can cause active over-barrier injection.

**3. Hot plasmonic carriers in confined systems.**

The most interesting case is a confined metal nanocrystal (**NC**) with localized electrons and plasmons. To describe such case, we will employ a method that combines a quantum approach for electron excitations and a classical approach for calculation of plasmonic fields. A NC model with infinite walls is shown in Figure 3. It has a shape of platelet/rectangular prism and can be submerged in water or located on/inside a crystal. Both cases are very important. The case of NC in a liquid solution is important for photo-chemistry and catalysis assisted by plasmonic electrons[5,6,7,9]. The geometry of a plasmonic prism (nano-antenna) is employed in opto-electronic photo-detectors[16,17,23].

First of all, we derive convenient equations for the electron Fermi gas in this system. Single-particle wave functions have the obvious form:

$$\psi_{\mathbf{n}=(n_x,n_y,n_z)} = \sqrt{\frac{2^3}{L_x L_y L_z}} Sin[k_{n_x} x] \cdot Sin[k_{n_y} y] \cdot Sin[k_{n_z} z],$$

$$\mathbf{k}_\mathbf{n} = \left(k_{n_x}, k_{n_y}, k_{n_z}\right) = \left(\frac{\pi}{L_x} n_x, \frac{\pi}{L_y} n_y, \frac{\pi}{L_z} n_z\right), \quad \varepsilon_\mathbf{n} = \frac{\hbar^2 \mathbf{k}_\mathbf{n}^2}{2m_0},$$

where $n_\alpha = 1, 2, 3...$ and the index $\alpha = x, y, z$. We now focus on an elongated NC with a squared base ($L_x = L_y < L_z$) and introduce convenient notations: $N_e$ is the total number of electrons in the Fermi gas and $n_F^{(\alpha)}$ is the maximum quantum number of electrons for the $\alpha$-direction, defined as $(\pi/L_\alpha) n_F^{(\alpha)} = k_F$. Then, the Fermi energy and the electron number are related through the following equations:

$$N_e = \frac{\pi}{3} n_F^{(x)} n_F^{(y)} n_F^{(z)},$$

$$n_F^{(x)} = n_F^{(y)} = \sqrt{E_F \frac{2mL_x^2}{\pi^2 \hbar^2}}, \quad n_F^{(z)} = \sqrt{E_F \frac{2mL_z^2}{\pi^2 \hbar^2}},$$

$$E_F = \frac{\hbar^2 k_F^2}{2m} = \frac{\hbar^2}{2m}\left(3\pi^2 \frac{N_e}{V_{NC}}\right)^{2/3}, \quad k_F = \frac{\pi}{L_x} n_F^{(x)} = \frac{\pi}{L_z} n_F^{(z)},$$

where $V_{NC} = L_z L_x^2$. Correspondingly the total number of electrons and the Fermi energy are connected by

$$N_e = \frac{V_{NC}}{3\pi^2}\left(\frac{2mE_F}{\hbar^2}\right)^{3/2}.$$

The hot carrier distributions are given by Eqs. 4 and can be rewritten conveniently in the form:

$$\delta\rho_{nn} = \frac{2e^2}{\Gamma} \sum_{\substack{\mathbf{n}=(n_x,n_y,n_z) \\ \mathbf{n'}=(n_x',n_y',n_z')}} (f_{\mathbf{n'}}^0 - f_{\mathbf{n}}^0)\left[|\varphi_{nn'}|^2 \frac{\Gamma}{(\hbar\omega - \varepsilon_n + \varepsilon_{n'})^2 + \Gamma^2} + |\varphi_{nn'}|^2 \frac{\Gamma}{(\hbar\omega + \varepsilon_n - \varepsilon_{n'})^2 + \Gamma^2}\right].$$

(6)

The matrix elements $\varphi_{nn'} = \langle n|\varphi_\omega(r)|n'\rangle$ include the potential $\varphi_\omega(r)$ which should be calculated numerically for a NC using the classical method of the local dielectric function. It is known that this method gives reliable results for metal NCs with dimensions above a few nanometers. For our NCs with sizes >5 nm, this approach can be applied. A cubic or platelet-like NC has no exact solution for the electric fields because of a complex geometry and we will use a numerical DDA method[44,45,46,47] which was successfully employed in a large number of studies. In the following, we employ the notations: $\varepsilon_0$ and $\varepsilon_{Au}$ will be the optical dielectric functions of matrix and metal, respectively. The matrix materials in this paper will be water ($\varepsilon_0 = 1.8$) and Si ($\varepsilon_0 = 13.7$). For a NC material, we will use gold and the data for $\varepsilon_{Au}$ from Ref.[48].

The optical matrix elements can be conveniently written in terms of the electric field:

$$\varphi_{nn',a} = \langle n|\varphi_\omega(r)|n'\rangle = \frac{\hbar^2}{m(\varepsilon_\mathbf{n} - \varepsilon_{\mathbf{n}'})} \int dV \cdot \psi_\mathbf{n} \left( \mathbf{E}_\omega \cdot \vec{\nabla} \psi_{\mathbf{n}'} \right), \quad (7)$$

where $\mathbf{E}_\omega = -\vec{\nabla}\varphi_\omega$ is the field inside a NC. A simplification for the above equation (7) comes from the fact that the electron transitions mostly appear at the Fermi surface with large numbers $|\mathbf{n}| \sim 20-40$ in a cubic gold NC with sizes $L_{NC} \sim 5-10 nm$. The number of electrons in such NCs $\sim 7 \cdot 10^3 - 6 \cdot 10^4$. Since the transitions mostly occur at the Fermi surface, the typical numbers $n$ and $n'$ are relatively large and the matrix element (7) rapidly decay with the difference $|\mathbf{n}-\mathbf{n}'|$. We now illustrate it with a simple example. Assume that the electric field $\mathbf{E}_\omega$ is constant and along the z-direction. Then, the element (7) becomes:

$$\varphi_{nn',a} = L_z E_\omega \cdot \frac{2}{\pi^2} \left( \frac{1}{(n_z'-n_z)^2} - \frac{1}{(n_z'+n_z)^2} \right) \delta_{n_x,n_x'} \delta_{n_y,n_y'} \approx L_z E_\omega \cdot \frac{2}{\pi^2} \frac{1}{(n_z'-n_z)^2} \delta_{n_x,n_x'} \delta_{n_y,n_y'}, \quad (8)$$

where $n-n'=odd$ because of the symmetry. This simple result shows an important property of the optical transitions: The squared matrix elements decrease very rapidly with increase of the difference $|\mathbf{n}-\mathbf{n}'|$; $|\varphi_{n'n,a}|^2 \sim |\mathbf{n}-\mathbf{n}'|^{-4}$ for the constant field inside a NC. For exact electric fields computed from the DDA program, we also observed a fast decrease of $|\varphi_{nm,a}|$ with increasing the difference $|\mathbf{n}-\mathbf{n}'|$. This tells us that important electronic excitations in a plasmon wave are in the vicinity of the Fermi level and the terms with relatively small $|\mathbf{n}-\mathbf{n}'|$ can be sufficient to describe accurately the spectrum of hot carriers.

The expression for the energy distribution of hot electrons can be conveniently rewritten in the form:

$$\delta n(\varepsilon) = 2 \sum_{\mathbf{n}=(n_x,n_y,n_z)} \delta\rho_{nn} \cdot \Phi(\varepsilon - \varepsilon_n),$$

where the coefficient 2 comes from the spins and $\Phi(\varepsilon - \varepsilon_n)$ plays a role of a delta function in the discrete sum:

$$\Phi(\varepsilon_0) = \frac{1}{\delta\varepsilon}, \quad |\varepsilon_0| < \delta\varepsilon/2,$$
$$\Phi(\varepsilon_0) = 0, \quad |\varepsilon_0| > \delta\varepsilon/2,$$

where $\delta\varepsilon$ should be taken as a small energy interval. Using Eq. 6, we write:

$$\delta n(\varepsilon) = 2e^2 \cdot 2 \sum_{\substack{\Delta \mathbf{n}=\mathbf{n}-(1,1,1) \\ \Delta \mathbf{n} \leq \mathbf{n}-(1,1,1)}} \sum_{\substack{\mathbf{n}=(n_x,n_y,n_z) \\ \mathbf{n} \geq (1,1,1)}} (f^0_{\mathbf{n}-\Delta\mathbf{n}} - f^0_{\mathbf{n}}) |\varphi_{\mathbf{n}-\Delta\mathbf{n},\mathbf{n},a}|^2 F_{\mathbf{n},\mathbf{n}-\Delta\mathbf{n}} \Phi(\varepsilon - \varepsilon_n) = \sum_{\substack{\Delta\mathbf{n}=\mathbf{n}-(1,1,1) \\ \Delta\mathbf{n}\leq\mathbf{n}-(1,1,1)}} \delta\rho^{(\Delta\mathbf{n})}(\varepsilon),$$

$$\delta n^{(\Delta\mathbf{n})}(\varepsilon) = 2e^2 \cdot 2 \sum_{\substack{\mathbf{n}=(n_x,n_y,n_z) \\ \mathbf{n}\geq(1,1,1)}} (f^0_{\mathbf{n}-\Delta\mathbf{n}} - f^0_{\mathbf{n}}) |\varphi_{\mathbf{n}-\Delta\mathbf{n},\mathbf{n},a}|^2 F_{\mathbf{n},\mathbf{n}-\Delta\mathbf{n}} \Phi(\varepsilon - \varepsilon_n),$$

$$F_{\mathbf{n},\mathbf{n}-\Delta\mathbf{n}} = \frac{1}{(\hbar\omega - \varepsilon_{\mathbf{n}} + \varepsilon_{\mathbf{n}-\Delta\mathbf{n}})^2 + \Gamma^2} + \frac{1}{(\hbar\omega + \varepsilon_{\mathbf{n}} - \varepsilon_{\mathbf{n}-\Delta\mathbf{n}})^2 + \Gamma^2}.$$

(9)

As we mentioned above, the matrix elements $|\varphi_{nn',a}|^2$ decrease very fast with the difference $\Delta n = n - n'$ and, therefore, the function $\delta\rho^{(\Delta\mathbf{n})}(\varepsilon_0)$ should also decay rapidly with $\Delta n$. In fact, the majority of hot electrons will be found in the several first terms $\Delta n$. Since we deal with relatively large NCs (such that $N_{NC} \gg 1$) the summation in the important equation (9) can be changed to the integral:

$$\delta n^{(\Delta\mathbf{n})}(\varepsilon_0) = 2e^2 \cdot 2 \int_{(n_x,n_y,n_z)>(0,0,0)} d^3\mathbf{n} \cdot (f^0_{\mathbf{n}-\Delta\mathbf{n}} - f^0_{\mathbf{n}}) \cdot |\varphi_{\mathbf{n}-\Delta\mathbf{n},\mathbf{n},a}|^2 F_{\mathbf{n},\mathbf{n}-\Delta\mathbf{n}} \delta(\varepsilon_0 - \varepsilon_n)$$
$$\approx 2e^2 |\varphi_{\Delta\mathbf{n},a}|^2 \cdot 2 \int_{(n_x,n_y,n_z)>(0,0,0)} d^3\mathbf{n} \cdot (f^0_{\mathbf{n}-\Delta\mathbf{n}} - f^0_{\mathbf{n}}) \cdot F_{\mathbf{n},\mathbf{n}-\Delta\mathbf{n}} \delta(\varepsilon_0 - \varepsilon_n)$$

(10)

In this equation, we assumed that $\varphi_{\mathbf{n}-\Delta\mathbf{n},\mathbf{n},a} \approx \varphi_{\Delta\mathbf{n},a}$ and $\Phi(\varepsilon - \varepsilon_n) \to \delta(\varepsilon - \varepsilon_n)$.

## 4. Nanocrystals in a solution for plasmonic chemical reactions.

We now consider particular cases of plasmonic nanocrystals in water and start with analytically solvable case of a platelet.

### 4.1 Plasmonic platelet.

An infinite slab of metal (Fig. 3a) has a simple solution for the electric fields. We now consider the case when the external electric field is perpendicular to the plate, i.e. $\mathbf{E}_0 \parallel \hat{\mathbf{z}}$, where $\mathbf{E}_0$ and $\hat{\mathbf{z}}$ are the external field and the normal to the plate, respectively (Fig. 4a). The field inside the platelet is easy to find from the boundary conditions:

$$\mathbf{E}_\omega = \gamma_{platelet}(\omega) \cdot \mathbf{E}_0, \quad \gamma_{platelet}(\omega) = \frac{\varepsilon_0}{\varepsilon_{Au}},$$

where $\mathbf{E}_\omega$ is the field inside the plate, and $\varepsilon_{Au}$ and $\varepsilon_0$ are the dielectric constants of gold and water, respectively. The dissipation of energy in a platelet is given by

$$Q_{NP} = \left\langle \int_{V_{NC}} dV\, \vec{j} \cdot \vec{E} \right\rangle_{time} = \mathrm{Im}(\varepsilon_{NP}) \frac{\omega}{2\pi} \int_{V_{NC}} dV \cdot \vec{E}_\omega \vec{E}_\omega^*,$$

where the current in a NC $\vec{j}_\omega = -i\omega(\varepsilon_{NP} - 1)/4\pi \cdot \vec{E}_\omega^{in}$. Correspondingly, the absorption cross section of a platelet is

$$\sigma_{NC} = \frac{Q_{NC}}{I_0} = \frac{\omega}{c_0\sqrt{\varepsilon_0}} V_{NC} \left| \frac{\varepsilon_0}{\varepsilon_{Au}} \right|^2 \mathrm{Im}(\varepsilon_{NP}),$$

where $I_0 = \frac{c_0\sqrt{\varepsilon_0}}{2\pi} \cdot E_0^2$ is the incident-light intensity. Figures 4c) and d) show the numerical result for the absorption cross-section of a platelet.

Since the field in the platelet is constant, the optical matrix elements $\varphi_{nn',a}$ can be analytically evaluated in the form of Eq. 8. Then, we should involve integrals given by Eq. 10. In the field configuration $\mathbf{E}_0 \| \hat{\mathbf{z}}$, only the number $n_z$ becomes changed, i.e. $\Delta\mathbf{n} = (0,0,\Delta n_z)$. Consequently, the hot-electron distribution is given by the sum over positive and odd $\Delta n_z$:

$$\delta n(\varepsilon) = \sum_{\substack{\Delta n_x = \Delta n_y = 0 \\ \Delta n_z = 1,3,5,...}} \delta n^{(\Delta n_z)}(\varepsilon),$$

$$\delta n^{(\Delta n_z)}(\varepsilon) = 2e^2 \left|\overline{\varphi}_{\Delta n_z}\right|^2 \cdot 2_{spin} \int_{(n_x,n_y,n_z)>(0,0,0)} d^3\mathbf{n}(f^0_{\mathbf{n}-\Delta\mathbf{n}} - f^0_{\mathbf{n}}) F_{\mathbf{n},\mathbf{n}-\Delta\mathbf{n}} \delta(\varepsilon - \varepsilon_n),$$

$$F_{\mathbf{n},\mathbf{n}-\Delta\mathbf{n}} = \frac{1}{(\hbar\omega - \varepsilon_{\mathbf{n}} + \varepsilon_{\mathbf{n}-\Delta\mathbf{n}})^2 + \Gamma^2} + \frac{1}{(\hbar\omega + \varepsilon_{\mathbf{n}} - \varepsilon_{\mathbf{n}-\Delta\mathbf{n}})^2 + \Gamma^2},$$

$$\overline{\varphi}_{\Delta n_z} = E_0 \cdot \gamma_{platelet}(\omega) \cdot \frac{2 L_z}{\pi^2} \frac{1}{\Delta n_z^2}, \quad \Delta n_z = 1,3,5,...$$

(11)

An important parameter entering Eq. 11 is the characteristic momentum transfer from a NC to an electron in the presence of an uniform excitation field. This characteristic momentum transfer is

$$q_L = \frac{\pi}{L_z}.$$

This momentum is involved in the optical transition $k_{n_z} \to k_{n_z + \Delta n_z}$, where $k_{n_z} = q_L n_z$. We will see soon that the characteristic momentum transfer $q_L$ is the key parameter of the problem. The allowed momentum transfers in the system are

$$q_{\Delta n_z} = q_L \cdot \Delta n_z,$$

where $\Delta n_z = odd$.

The integral in Eq. 11 can be taken analytically (see Supporting Information):

$$\delta n^{(\Delta n_z)}(\varepsilon) = 4 \cdot e^2 \left| \overline{\varphi}_{\Delta n_z} \right|^2 \cdot W^{(\Delta n_z)}(\varepsilon),$$

$$W^{(\Delta n_z)}(\varepsilon_0) = -\frac{1}{2\Gamma} \left( \frac{V}{L_z^3} \right) \frac{\pi}{\left(E_L^2 \Delta n_z\right)} \times$$

$$\times \left[ AcrTan\left(\frac{x_c - x_0}{\gamma}\right) - AcrTan\left(\frac{1 - x_0}{\gamma}\right) + AcrTan\left(\frac{x_c + x_1}{\gamma}\right) - AcrTan\left(\frac{1 + x_1}{\gamma}\right) \right], \quad (12)$$

$$E_L = \frac{\hbar^2 \pi^2}{mL_z^2}, \quad n_0 = \sqrt{\frac{2\varepsilon_0}{E_L}}, \quad x_c = Cos[\theta_c] = \frac{n_0^2 - n_F^2 + \Delta n_z^2}{2\Delta n_z \cdot n_0}, \quad n_F = \sqrt{E_F \frac{2mL_z^2}{\pi^2 \hbar^2}},$$

$$x_0 = \frac{\hbar\omega}{E_L \Delta n_z \cdot n_0} + \frac{\Delta n}{2n_0}, \quad x_1 = \frac{\hbar\omega}{E_L \Delta n_z \cdot n_0} - \frac{\Delta n}{2n_0}, \quad \gamma = \frac{\Gamma}{E_L \Delta n_z \cdot n_0}.$$

This complicated equation can be simplified for an important case of small $\Delta n_z$ when the energy of hot electrons is low $\varepsilon_\mathbf{n} - \varepsilon_{\mathbf{n}-\Delta\mathbf{n}} \sim \Delta n \cdot \hbar q_L v_F \ll \hbar\omega$. Then, this function becomes

$$\delta n^{(\Delta n_z)}(\varepsilon) = 4 \cdot e^2 \left| \overline{\varphi}_{\Delta n_z} \right|^2 \left( \frac{V_{NC}}{L_z^3} \right) \frac{n_F^{(z)}}{(\hbar\omega)^2 E_L \pi} \frac{E_F - \varepsilon + \hbar v_F q_{\Delta n}}{\hbar v_F q_{\Delta n}}, \quad n_F^{(z)} = \sqrt{E_F \frac{2mL_z^2}{\pi^2 \hbar^2}}, \quad E_F < \varepsilon < E_F + \hbar v_F q_{\Delta n}.$$
(13)

In this case, when $\varepsilon_\mathbf{n} - \varepsilon_{\mathbf{n}-\Delta\mathbf{n}} \ll \hbar\omega$, the excitation of electrons happen without direct conservation of energy like in a 3D plasmon wave in bulk (see Section 2 above). The hot-energy distribution has a triangular shape and occupies the interval $E_F < \varepsilon < E_F + \hbar v_F q_L \cdot \Delta n_z$. This case resembles a 3D plasmon wave. For large $\Delta n$, when $\varepsilon_\mathbf{n} - \varepsilon_{\mathbf{n}-\Delta\mathbf{n}} \sim \hbar\omega$, the function $\delta n^{(\Delta n_z)}(\varepsilon)$ is very different and depends strongly on the relaxation rate $\Gamma$. The main contribution in this case comes from the pole-like behavior in the function $F_{\mathbf{n},\mathbf{n}-\Delta\mathbf{n}}$. This pole-like behavior appears when the conservation of energy and momentum happen simultaneously (i.e. the case $\varepsilon_{\mathbf{n}} - \varepsilon_{\mathbf{n}'-\Delta\mathbf{n}} = \hbar\omega$). In this case,

$$\delta n^{(\Delta n_z)}(\varepsilon) = 2 \cdot e^2 \frac{\left|\bar{\varphi}_{\Delta n_z}\right|^2}{\Gamma} \cdot \frac{V_{NC}}{L_z^3} \frac{\pi^2}{E_L^2 \Delta n_z}. \qquad (14)$$

This term $\delta n^{(\Delta n_z)}(\varepsilon)$ does not depend on the energy. We also can see that $\delta n^{(\Delta n)} \sim 1/\Gamma$. Such contributions appear when the electron is directly excited by a photon under the condition $\varepsilon_{\mathbf{n}} - \varepsilon_{\mathbf{n}'-\Delta\mathbf{n}} = \hbar\omega$. We note that the condition $\varepsilon_{\mathbf{n}} - \varepsilon_{\mathbf{n}'-\Delta\mathbf{n}} = \hbar\omega$ does not appear for small $\Delta n_z$ in NCs with realistic sizes ~ 5nm or larger. For the energy relaxation constant, we will take a typical number known from the time-resolved spectroscopy of gold NCs, $\Gamma = \hbar/\tau = 1.3 meV$, where the relaxation time $\tau = 0.5 ps$ [49].

Numerical and analytical methods coming from Eqs. 9 and 10 give similar results (see Supporting Information). In Figure 4, we show the calculated hot-electron distributions for platelets with various widths. We can see very characteristic shapes. The triangular contribution at small energies $E_F < \varepsilon < E_F + \hbar v_F q_L$ comes from the term $\Delta n_z = 1$ (approximated by Eq. 13). The term $\Delta n_z = 3$ is typically suppressed because of the matrix element $\left|\bar{\varphi}_{\Delta n_z}\right| \propto \Delta n_z^{-4}$. But, for the numbers $\Delta n_z \geq \Delta n_{critical} = \hbar\omega/(\hbar q_L v_F)$, the terms $\delta n^{(\Delta n)}(\varepsilon)$ may become again large or noticeable and originate from the transitions in the confined plasma with conservation of energy when $\varepsilon_{\mathbf{n}} - \varepsilon_{\mathbf{n}'-\Delta\mathbf{n}} = \hbar\omega$. These terms give flat contributions in the maximally-possible energy interval $E_F < \varepsilon < E_F + \hbar\omega$. It is important to look at the electron distributions $\delta n(\varepsilon)$ as a function of the NC width $L_z$.

For small $L_z \sim 5-10 nm$, the hot electrons occupy substantially the whole allowed interval $E_F < \varepsilon < E_F + \hbar\omega$. This regime for gold corresponds to the numbers $\Delta n_{critical} = \hbar\omega / (\hbar q_L v_F) \sim 3-10$. But, for $L_z > 15 nm$, the hot-electron distribution becomes strongly peaked in a very narrow region of small energies, $E_F < \varepsilon < E_F + \hbar v_F q_L$. It is especially well seen if we normalize the electron distribution to the extinction or to the volume (extinction $\sigma_{NC} \sim V_{NC}$). We see this important behavior in Figs. 4 and S2 for two types of NC, a cube and a platelet. The sketch in Figure 4a shows the characteristic energies of excitations in NCs. In small NCs, hot electrons are distributed in the interval $E_F < \varepsilon < E_F + \hbar\omega$, whereas in larger NCs hot electrons are concentrated in the narrow region $E_F < \varepsilon < E_F + \hbar v_F q_L$. For photo-induced chemical reactions and devices based on hot-electron injection, it is preferable to have a plasmonic nano-antenna with a small width (<20nm) and to illuminate a NC with light polarized along a small-size direction (i.e. along the z-direction in our case).

We now look at the numbers of photo-excited electrons of a platelet in the two intervals: $E_F < \varepsilon < E_F + \hbar v_F q_L$ and $E_F + \hbar v_F q_L < \varepsilon < E_F + \hbar\omega$. Performing integration over energy, we obtain:

$$\delta N_{low-energy} = \int_{E_F}^{E_F + \hbar v_F q_{NC}} d\varepsilon \delta n(\varepsilon) \sim \frac{|\gamma_{platelet}(\omega)|^2}{(\hbar\omega)^2} \left(\frac{V_{NC}}{L_z^3}\right) L_z^4 \sim L_z^2,$$

$$\delta N_{high-energy} = \int_{E_F + \hbar v_F q_{NC}}^{E_F + \hbar\omega} d\varepsilon \delta n(\varepsilon) \sim \frac{|\gamma_{platelet}(\omega)|^2}{\Gamma(\hbar\omega)^4} \left(\frac{V}{L_z^3}\right) L_z^2 \sim L_z^0,$$

(15)

where $V_{NC} = L_z L_x^2$ is the NC volume, as before. In the above equations (15), we included only important factors related to the NC size and the frequency. To make the estimates (15), we did the following. For $\delta N_{low-energy}$, we included only the first term $\delta n^{(\Delta n_z=1)}$. To get an estimate for $\delta N_{high-energy}$, we approximated a sum over $\Delta n_z$ as an integral:

$$\sum_{\Delta n_z = \Delta n_{crit},...} \Delta n_z^{-5} \sim \int_{\Delta n_{crit}} x^{-5} dx \sim \Delta n_{crit}^{-4} \sim (L_z \omega)^{-4}.$$

Importantly, the ratio between the above numbers increases rapidly with the NC size

$$\frac{\delta N_{low-energy}}{\delta N_{high-energy}} \sim L_z^2.$$

In small NCs, the number of high-energy electrons in the interval $E_F + \hbar v_F q_L < \varepsilon < E_F + \hbar \omega$ is larger and this case is more suitable for applications of hot carriers. In large NCs, most of the hot electrons have small energies and these energies are in the interval $E_F < \varepsilon < E_F + \hbar v_F q_L$, which is very close to the Fermi level. In Supporting Information, we plot explicitly the ratio $\delta N_{low-energy} / \delta N_{high-energy}$ for the case of plasmonic cube (Figure S3).

### 4.2 Plasmonic cube in solution

Plasmonic cube in water can also be solved with this formalism. The difference is in the numerical solution for the internal field. Using calculated internal fields and Eqs. 9 and 10, one can find the matrix elements and then the hot-electron distributions. Figure 5 now shows the results for the Au cube. The electric fields inside a cube are enhanced stronger compared to a platelet and sphere and, importantly, the plasmon peak is shifted to the red. In this way, one can tailor an optical response to make better spectral overlap with the solar spectrum, for example. In Supporting Information (Figure S4), we also show a comparison of the fields between an Au cube and a sphere. An averaged electric field in a NC is defined as

$$\bar{E}_{NC} = \left( \int_{V_{NC}} \frac{\vec{E}_\omega \vec{E}_\omega^*}{E_0^2} dV \right)^{1/2}.$$

We can see that, in general, the fields in a cube and in a sphere are similar. Therefore, for estimates in cubic NCs, one can use analytical results for a sphere. In a nano-sphere, the internal electric field is given by a simple equation:

$$\mathbf{E}_\omega = \gamma(\omega) \cdot \mathbf{E}_0, \ \gamma(\omega) = \frac{3\varepsilon_0}{\varepsilon_0 + 2\varepsilon_{Au}},$$

where $\mathbf{E}_0$ is the incident field. In other words, one can approximate a field in a nano-cube as an uniform field inside a sphere. This gives a simple way to describe semi-quantitatively the properties of cubic NCs.

**4.3 Rate of injection from plasmonic nanocrystals for photo-chemical reactions**

Now we turn to discussion on possible applications of hot electrons for photo-chemistry[5,9,12,13]. Figure 5a illustrates transfer of photo-generated carriers to surface molecules. In this case, the key parameter is the number of photo-excited carriers at a given energy $\varepsilon_{mol}$ that corresponds to the state of charged molecule (negatively-charged for a reduced molecule and positively-charged for an oxidized one). The number of transferred hot electrons per second is proportional to the hot electron (hole) population at the energy of a molecule:

$$Rate_{e(h)} = \gamma_{tun} \cdot \frac{\delta n_{e(h)}(\varepsilon_{mol})}{n_{DOS}(\varepsilon_{mol})},$$

where $\varepsilon_{mol}$ is the energy of a reacting molecule (Fig. 5a), $\gamma_{tun}$ is the tunneling rate, and $n_{DOS}(\varepsilon_{mol})$ is the density of states in a NC at the energy $\varepsilon_{mol}$. The tunnel rate can be written as $\gamma_{tun} = 2\pi \cdot w_t^2 \cdot \rho_{DOS}(\varepsilon_{mol})$ (See Supporting Information). Another interesting parameter is an efficiency of generation of photo-carriers

$$Eff(\omega) = \frac{Rate_{e(h)}(\omega)}{(\sigma_{NC} I_0 / \hbar\omega)} \propto \frac{\delta n_{e(h)}(\varepsilon_{mol}, \omega, L_{NC})}{(\sigma_{NC} I_0 / \hbar\omega)}.$$

Figure 5b presents the efficiency as a function of the size of an Au cube at the plasmon resonance frequency ($\hbar\omega = 2.22 eV$) which is in the green region of the solar spectrum. The energy of negatively-changed state of molecule was taken in the form $\varepsilon_{mol} = \varepsilon_F + \Delta\varepsilon_{mol}$ with $\Delta\varepsilon_{mol} = 0.5$ and $1.9 eV$. We see that the efficiency rapidly decreases with a size of a NC. This is due to the parameter $q_L = \pi / L_z$ which determents efficiency of momentum transfer from a NC to mobile electrons. In other words, for larger $q_L = \pi / L_z$ (or smaller $L_z$) non-conservation of electron momentum in a NC is more efficient and more high-energy electrons become generated. This property can also be formulated in terms of a characteristic energy $\hbar q_L v_F$. Our numerical results show that efficient generation of hot electrons with energies $\sim E_F + \hbar\omega$ appears when the parameter $\Delta n_{crit} = \hbar\omega / \hbar q_L v_F$ is not very large, ~ *5-10* or $L_z \sim 5-15nm$. In Fig. 5b we also give numbers for the parameter $\Delta n_{crit}$. The efficiency in Fig. 5b shows oscillations due to the quantum-size effect in a NC. Such oscillations are not seen typically in experiments since NCs in solution have a dispersion of sizes and also experience thermal fluctuations.

To conclude this section, we consider the ratio $\delta N_{low-energy} / \delta N_{high-energy}$ for a cube as a function of a NC size (Fig. S3b). We see again that the high-energy electrons dominate in the hot-electron spectra of small NCs and, for larger NCs, the majority of electrons have small energies. Figure S3c also shows the rate of generation of hot carriers in an Au- cube defined as

$$Rate_{hot\ electrons} = \frac{\Gamma}{\hbar}\delta N_{tot},$$

$$\delta N_{tot} = \int_{E_F}^{\infty} d\varepsilon \cdot \delta n(\varepsilon).$$

The function $Rate_{hot\ electrons}(\omega)$, of course, shows a plasmon resonance. However, the calculated rate does not include the inter-band transitions in Au and, therefore, the function $Rate_{hot\ electrons}(\omega)$ does not have the near-UV spectral structure at 300-400nm that is strong in the absorption spectrum of Au (Fig. S3c).

**5. Injection of hot plasmonic electrons from a metal slab into a semiconductor contact.**

When a nano-antenna (platelet) or a 2D slab is located on a semiconductor substrate, hot-plasmonic electrons can be injected into a semiconductor (Figures 6 and 7). Schottky barrier efficiently traps transferred electrons and this leads to a photo-current[1,16].

In Figure 6, we chose the geometries in which the polarization of incident light is perpendicular to the plane of nano-antenna. This is the most preferable configuration for generation of high-energy electrons (with energies $\sim \hbar\omega$) in a NC. We now consider injection from plasmonic slabs. In this case, the important factor for the injection process is conservation of the electron momentum parallel to the boundary[22]. This factor limits the number of excited electrons which can enter the contact. In fact, only excited electrons within a cone in the momentum space can move above the barrier[22]. The condition for an electron to move through a Schottky barrier is

$$\varepsilon_z = \frac{p_z^2}{2m} > E_B \text{ or } n_z > n_B,$$

where $E_B$ is the barrier energy (Figure 7a) and the critical z-quantum number of electron ($n_B$) is given by the equation

$$E_B = \frac{\hbar^2 \pi^2 n_B^2}{2m L_z^2}.$$

Similarly, the critical z-component of electron wavevector for transfer to the contact is $k_{z,B} = k_B = q_L n_B$, where

$$n_B = \sqrt{\frac{2m L_z^2 E_B}{\hbar^2 \pi^2}}.$$

Another important parameter is a barrier height relative to the Fermi energy

$$\Delta E_B = E_B - E_F.$$

Figure 7a illustrates the transfer process and shows the energy diagram.

Then, the total number of electrons, which are allowed to move into a contact, is given by an integral over the population of all quantum states with $n_z > n_B$ and arbitrary $n_x$ and $n_y$:

$$\delta n_{transfer}(\varepsilon) = \sum_{\substack{\Delta n_x = \Delta n_y = 0 \\ \Delta n_z = 1,3,5,\ldots}} \delta n_{transfer}^{(\Delta n_z)}(\varepsilon),$$

$$\delta n_{transfer}^{(\Delta n_z)}(\varepsilon) = 2e^2 \left| \overline{\varphi}_{\Delta n_z} \right|^2 \cdot 2 \int_{(n_x,n_y,n_z) > (0,0,n_B)} d^3\mathbf{n}(f_{\mathbf{n}-\Delta\mathbf{n}}^0 - f_{\mathbf{n}}^0) F_{\mathbf{n},\mathbf{n}-\Delta\mathbf{n}} \delta(\varepsilon - \varepsilon_n),$$

$$F_{\mathbf{n},\mathbf{n}-\Delta\mathbf{n}} = \frac{1}{\left(\hbar\omega - \varepsilon_{\mathbf{n}} + \varepsilon_{\mathbf{n}-\Delta\mathbf{n}}\right)^2 + \Gamma^2} + \frac{1}{\left(\hbar\omega + \varepsilon_{\mathbf{n}} - \varepsilon_{\mathbf{n}-\Delta\mathbf{n}}\right)^2 + \Gamma^2},$$

$$\overline{\varphi}_{\Delta n_z} = E_0 \cdot \gamma_{platelet}(\omega) \cdot \frac{2L_z}{\pi^2} \frac{1}{\Delta n_z^2}, \quad \Delta n_z = 1,3,5,\ldots$$

(16)

This equation differs from Eq. 11 only by the limits of integration. The integrals in (16) can again be taken analytically (see Supporting Information) and are written here:

$$\delta n_{transfer}^{(\Delta n_z)}(\varepsilon) = 4 \cdot e^2 \left| \overline{\varphi}_{\Delta n_z} \right|^2 \cdot \tilde{W}^{(\Delta n_z)}(\varepsilon),$$

$$\tilde{W}^{(\Delta n_z)}(\varepsilon_0) = -\frac{1}{2\Gamma} \left(\frac{V}{L_z^3}\right) \frac{\pi}{\left(E_L^2 \Delta n_z\right)} \times$$

$$\times \left[ AcrTan\left(\frac{x_{B,c} - x_0}{\gamma}\right) - AcrTan\left(\frac{1 - x_0}{\gamma}\right) + AcrTan\left(\frac{x_c + x_1}{\gamma}\right) - AcrTan\left(\frac{1 + x_1}{\gamma}\right) \right],$$

$$E_L = \frac{\hbar^2 \pi^2}{mL_z^2}, \quad n_0 = \sqrt{\frac{2\varepsilon_0}{E_L}}, \quad x_{B,c} = \begin{cases} Cos[\theta_B] = \dfrac{n_B}{n_0}, & n_0 < n_{crit} \\ Cos[\theta_c] = \dfrac{n_0^2 - n_F^2 + \Delta n_z^2}{2\Delta n_z \cdot n_0}, & n_{max} = n_F + \Delta n > n_0 > n_{crit} \end{cases},$$

$$n_{crit} = \sqrt{2\Delta n \cdot n_B + n_F^2 - \Delta n^2}, \quad n_F = \sqrt{E_F \frac{2mL_z^2}{\pi^2 \hbar^2}},$$

$$x_0 = \frac{\hbar\omega}{E_L \Delta n_z \cdot n_0} + \frac{\Delta n}{2n_0}, \quad x_1 = \frac{\hbar\omega}{E_L \Delta n_z \cdot n_0} - \frac{\Delta n}{2n_0}, \quad \gamma = \frac{\Gamma}{E_L \Delta n_z \cdot n_0}.$$

(17)

We now introduce figures of merit for generation and injection of carriers in the structures shown in Fig. 6. First we define the integral numbers of photo-generated carriers in a NC:

$$\delta N_{tot} = \int_{E_F}^{E_F+\hbar\omega} \delta n(\varepsilon) \cdot d\varepsilon,$$

$$\delta N_{generated,\,\varepsilon>E_B} = \int_{E_F+E_B}^{E_F+\hbar\omega} \delta n(\varepsilon) \cdot d\varepsilon,$$

$$\delta N_{injected,\,k_z>k_B} = \int_{E_F+E_B}^{E_F+\hbar\omega} \delta n_{transfer}(\varepsilon) \cdot d\varepsilon,$$

where $\delta N_{tot}$ is the total number of excited electrons, as it was defined before, $\delta N_{generated,\,\varepsilon>E_B}$ is the number of excited electrons above the threshold energy of the barrier $E_B$, and $\delta N_{injected,\,k_z>k_B}$ is the number of electrons that can move into the contact. The rates of generation of carriers are given by multiplication of the above numbers by the relaxation rate $\gamma = \Gamma/\hbar$:

$$Rate_{tot} = \gamma \cdot \delta N_{tot},$$
$$Rate_{generated,\,\varepsilon>E_B} = \gamma \cdot \delta N_{generated,\,\varepsilon>E_B},$$
$$Rate_{injected,\,k_z>k_B} = \gamma \cdot \delta N_{injected,\,k_z>k_B}.$$

We note that the parameter $Rate_{injected,\,k_z>k_B}$ represents a maximum possible rate of injection assuming that all generated carriers with $k_z > k_B$ are transferred to a contact. The reflection of carriers at the metal-semiconductor interface can decrease injection. The rate $Rate_{generated,\,\varepsilon>E_B}$ is essential for cases when the parallel momentum (in the x-y plane) is not conserved or injection can appear from all interfaces (the case of a cube). In Figures 9 and 10, we plot such rates for an Au slab, platelet, and cube. The corresponding efficiencies can be defined in the following way. The electron efficiencies of generation and injection:

$$Eff_{\varepsilon > E_B} = \frac{Rate_{generated, \varepsilon > E_B}}{Rate_{tot}},$$

$$Eff_{injection} = \frac{Rate_{injected, k_z > k_B}}{Rate_{tot}}.$$

These efficiencies describe the fractions of generated (above $E_B$) and injected carriers from the total number of excited electrons. We now introduce a quantum efficiency of injection

$$QE_{injected} = \frac{Rate_{injection, k_z > k_B}}{Rate_{photon\ absorption}},$$

$$Rate_{photon\ absorption} = \frac{\sigma_{abs} I_0}{\hbar \omega},$$

where $Rate_{photon\ absorption}$ is the rate of absorption of photons and the cross section $\sigma_{abs}$ was defined above. Figures 10 and 11 present the results. In these figures, we also introduced an averaged-over-frequency-range quantum efficiency. Since carriers are created everywhere inside a NC, the number of injected carriers can be reduced due to a finite mean free path $l_{mfp}$. A probability to reach the interface for an electron $P(x) = e^{-x/l_{mfp}}$, where $x$ is a distance to travel. If electrons are created inside a slab, the averaged probability for injection through one interface will be

$$P_1 = \frac{1}{L_{NC}} \int_0^{L_{NC}} dx \cdot P(x) = \frac{l_{mfp}}{L_{NC}} (1 - e^{-L_{NC}/l_{mfp}}).$$

This equation assumes transfer of electrons only through one interface of a slab, like in Fig. 6a. If a slab has injection from both interfaces (Fig. 6b), the distance for an electron to travel to an interface is $\sim L_{NC}/2$ and the averaged probability is

$$P_2 = \frac{2 \cdot l_{mfp}}{L_{NC}}(1 - e^{-L_{NC}/2 \cdot l_{mfp}}).$$

Figures 10 and 11 also include plots corrected by the factors $P_{1(2)}$.

A general observations form the numerical results is that the efficiencies of generation and injection decrease with the width of NC. The decrease comes from two reasons. (1) The electron distribution in a large NC has majority of carries at low energies (Fig. 10a). These low-energy carriers cannot be injected. (2) A simple effect of scattering does not allow some of the electrons to move freely from an inside of NC to an interface. This effect can be described by incorporating a mean free path. According to our calculations, platelets and cubes with dimensions of about 10nm are most suitable for generation of carriers with high energy.

The injection rates calculated above for plasmonic slabs assumed ideal interfaces and conservation of the momentum parallel to the metal-semiconductor boundary. In real nanostructures, imperfections and roughness of the interfaces can remove this condition. We note that the conservation of parallel momentum for ballistic electrons is a limiting factor for the injection current. In Figures 8b and 9a, we can see that the conservation of parallel momentum leads to a strong decrease of the injection rate. Therefore, one can speculate that interface imperfections may enhance injection currents since they break the parallel-momentum conservation.

## 6. Comparison with the Fowler theory.

The Fowler law[1,22] for the injection from a metal to a semiconductor is remarkably simple and general

$$I_{photo-current} = I_0 \cdot (\hbar\omega - \Delta E_B)^2. \qquad (18)$$

The derivation by Fowler[22] is based on two assumptions. Hot electrons are generated with isotropic momentum distribution and only a fraction of electrons with high enough perpendicular velocity (the condition $\varepsilon_z = \frac{p_z^2}{2m} > E_B$) can move into the contact. These two conditions give the exponent 2 in Eq. 18. The paper by Fowler[22] also made attempt to derive the coefficient in (18) using qualitative arguments on the wave function. The time-dependent wave function was written for the moment of collision with the barrier in the presence of time-dependent electric field of light. Here we use a different approach and can estimate the current from the metal using the solution for the steady state of density matrix of a Fermi gas in a metal slab. The fraction of electrons that can enter the contact is given by the area shown in Fig. 7b. The number of photo-generated high-energy electrons per one quantum state in the momentum space can be estimated using Eqs. 12, 13, and 14. In Eq. 12, we need to perform summation over the terms $\Delta n_z$ assuming $\Delta n_z > \Delta n_{z,crit} \sim \hbar\omega / \hbar q_L v_F$. The sum converges rapidly and was already estimated above as an integral (see the derivation of Eqs. 15). The resulting analytical equation for the maximum possible injected current from a metal slab reads

$$I_{injected,barrier} = Rate_{injected, k_z > k_B} = P_b \cdot \gamma \cdot \delta N_{injected, k_z > k_B} = C_0 \left( \hbar\omega - \Delta E_B \right)^2,$$

$$C_0 = P_b \frac{2}{\hbar \pi^2} \frac{A \cdot |eE_{z,in}|^2}{E_B} \frac{E_F^2}{\Delta E_B^4},$$
(19)

where we have assumed $\hbar\omega \approx \Delta E_B$; $A$ is a surface area of a slab and $E_{z,in}$ is the z-component of the electric field inside the metal. For simplicity, we did not include here the effect of mean free path assuming that a slab is relatively narrow. In addition, we introduced a coefficient $P_b$ describing the probability for a photo-excited electron to be injected into a contact; this parameter also includes the quantum transparency of the metal-semiconductor contact for photo-generated electrons with $\varepsilon_z = \frac{p_z^2}{2m} > E_B$. For $P_b = 1$, all electrons with with $\varepsilon_z > E_B$ can enter the contact. In other words, Eq. 19 with $P_b = 1$ gives an estimate of the maximum possible injection rate. The obtained equation (19) is interesting from the following point of view. It does not depend on the width of a slab $L_{NC}$. The parameter $L_{NC}$ simply dropped out from the equation. It tells us that the derived equation describes a "surface" properly of our system - the inelastic optically-induced scattering of electron by a wall that leads to non-conservation of momentum. Non-conservation of momentum is needed to activate inter-level transitions under the action of light field. This momentum change leads to the electron transitions with conservation of energy which satisfy the condition $\varepsilon_\mathbf{n} - \varepsilon_{\mathbf{n}-\Delta\mathbf{n}} = \hbar\omega$. Equation (19) reproduces the numerical data in Fig. 8c. The numerical data in Fig. 8c show the size-quantization effect seen as discrete peaks. Equation (19) does not include the size-quantization effect, but it describes an envelope for the numerical curve in Fig. 8c. We note that our derivation for the injection current was done under the assumption of a closed system (NCs with hard walls), whereas the previous derivations of the photoelectric currents were obtained for a step-like barrier potential[29,30].

Equation (19) depends on the z-component of the external field. However some experiments have shown an anisotropic character of injection currents[50]. It can be that, in metal films, non-conservation of momentum can also come from defects and phonons and it may lead to the corresponding contribution to the photoelectric current. Another contribution can appear due to the inter-band transitions in a metal NC considered above. If so, the current may have three contributions

$$I_{\text{injected, tot}} = I_{\text{injected, barrier}} + I_{\text{injected, defects}} + I_{\text{injected, inter-band transitions}},$$

where $I_{\text{injected, defects}}$ is the current of hot electrons generated optically with a help of defects that break momentum conservation. This defect-induced photo-current can be isotropic with respect to optical excitation. The current $I_{\text{injected, inter-band transitions}}$ comes from the optical transition between the d- and sp-bands shown in Figure 2f. As it was discussed before in Section 2.3, the inter-band transitions become active at $\hbar\omega > 2.4 eV$. At the threshold of the inter-band transitions ($\hbar\omega \sim 2.4 eV$), hot plasmonic electrons have small excitation energies (i.e. they are generated in the vicinity of the Fermi level) and cannot be used for injection, whereas the photo-generated holes of the d-band are created far from the Fermi surface (with energies $\varepsilon_F - 2.4 eV$) and may be useful for injection.

## 7. Relaxation times of single hot electrons and plasmons.

In the approach of the quantum master equation (1) used above, the relaxation of energy of electrons is given by the parameter $\Gamma = \hbar/\tau$, where the energy-relaxation time $\tau = 0.5\,ps$ was taken from the experiment[49]. We note that our system has another relevant relaxation time – the time of coherent relaxation of a plasmon in a metal NC, $\tau_{plasmon} = \hbar/\Gamma_{plasmon}$. This time determines the width of the plasmon resonance ($\Gamma_{plasmon}$) and enters the dielectric function of metal which is used to calculate the field factor $\gamma(\omega)$. In the cases of platelet and sphere, the field factor has simple analytical expressions given above. For the platelet, $\gamma_{platelet}(\omega) \sim 1/\varepsilon_{Au}$ and $\delta n_{e(h)} \sim |\gamma_{platelet}(\omega)|^2 \sim |\varepsilon_{Au}|^{-2}$. The broadening of the plasmon resonance in the dielectric function, of course, strongly influences the strength of the generation of hot carriers. This broadening for a spherical NC can be written as [51,52,53]

$$\Gamma_{plasmon} = \Gamma_{bulk} + v_F/R,$$

where $\Gamma_{bulk}$ is the broadening in a bulk metal and the second terms appears due to the electron collisions with a surface in a NC with radius $R$. The plasmon decay time can be estimated from the bulk dielectric function of gold[44] that gives $\Gamma_{bulk} \sim 76\,meV$. Then, the calculated time turns out to be rather short $\tau_{plasmon} \sim 8\,fs$. The relaxation time in gold is short because of rapid de-phasing of plasmon oscillations due to collisions with phonons and defects.

## 8. Conclusions

To summarize, we developed here a quantum theory of generation of plasmonic hot electrons in metal NCs. Our approach is based on the random-phase approximation. We write down a steady-state solution for the density matrix and then integrate the distribution function in the momentum space. This gives simple and reliable results for the distributions of plasmonic carries in NCs. The quantum confinement plays a very important role providing a momentum to electron excitations. The key parameters of the problem are

$$\Delta n_{crit} = \frac{\hbar \omega}{\hbar q_L v_F} \text{ and } q_L = \frac{\pi}{L_{NC}}.$$

In gold nanocrystals, when $\Delta n_{crit} \sim 10$, the high-energy electron generation is efficient. For larger parameters $\Delta n_{crit}$, the number of high-energy electrons is greatly reduced. Our results show that efficient generation of plasmonic carriers with high excitation energies ( $\sim \hbar \omega$) occurs in NCs with small widths ~10-20nms. Nanostructures suggested by us as efficient electron injectors are slabs, platelets, cubes, nanowires with relatively-small widths ($L_z \sim 10-20nm$). These nano-antennas may have larger sizes in the other directions ($L_{x(y)} \sim 50-100nm$). Importantly, for efficient generation of plasmonic carriers, the polarization of electric field of incident light should be along the small dimension of a nano-antenna. Our results can also be applied for photo-activated electron transport in arrays of metal nanoparticles on substrates[54,55,56,57]. Transport in arrays of NCs can be thermally activated (1) or due to non-equilibrium plasmonic carriers (2). Our equations can be used to describe the mechanism (2) which involves high-energy plasmonic electrons under steady-state illumination. The thermal mechanism (1) can give a rather weak contribution for systems with efficient removal of heat. In addition to direct injection of carriers from plasmonic NC, plasmons can enhance absorption of a neighboring semiconductor nanostructure[58,59] and, in this way, amplify chemical processes or photo-currents[12]. The advantage of plasmonic NCs for hot-electron applications is in large absorption cross sections, whereas a challenge is in fast relaxation of hot carriers. This is in contrast to semiconductor NCs used intensively in experiments for photo-chemical reactions[60].

To conclude, our theory can be applied to a variety of chemical and physical systems. These systems include NC solutions with plasmon-assisted chemical reactions and solid-state plasmonic devices incorporating metal nano-antennas, semiconductor contacts, and Schottky barriers. Our results show that the size, design, and field directionality are crucial for efficient generation and injection of plasmonic carriers.

**Acknowledgements**. This work was supported by the Science Foundation Ireland and by the NSF (project: CBET-0933782). A.O.G. thanks Mark Knight for discussions.

**Supporting Information.** It includes details of derivations and additional numerical data. This information is available free of charge via the Internet at http://pubs.acs.org.

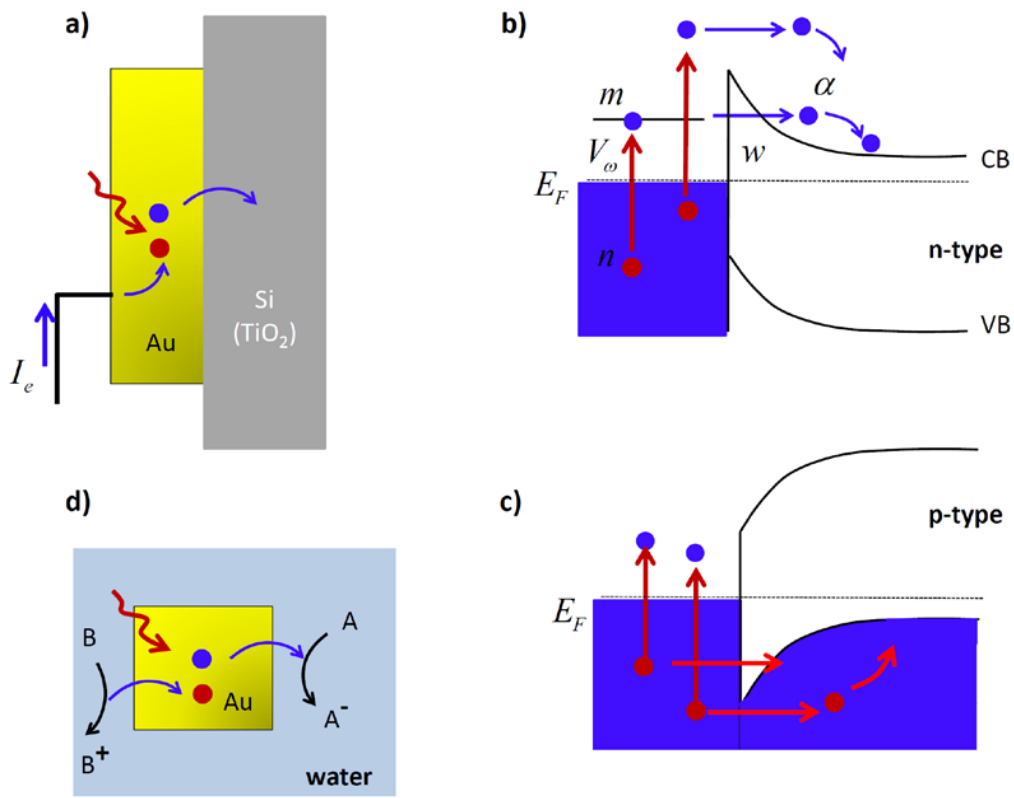

**Figure 1**. Models used in the text. The model **(a)** describes injection of carriers from a metal NC to a semiconductor. **b)** and **c)** Band diagrams of the metal-semiconductor junction with optical and transport processes. Transport processes include both tunneling and ballistic transport. The case **(d)** shows a schematics of nanocrystal in a solution; photo-generated electrons and holes can be transferred to molecules.

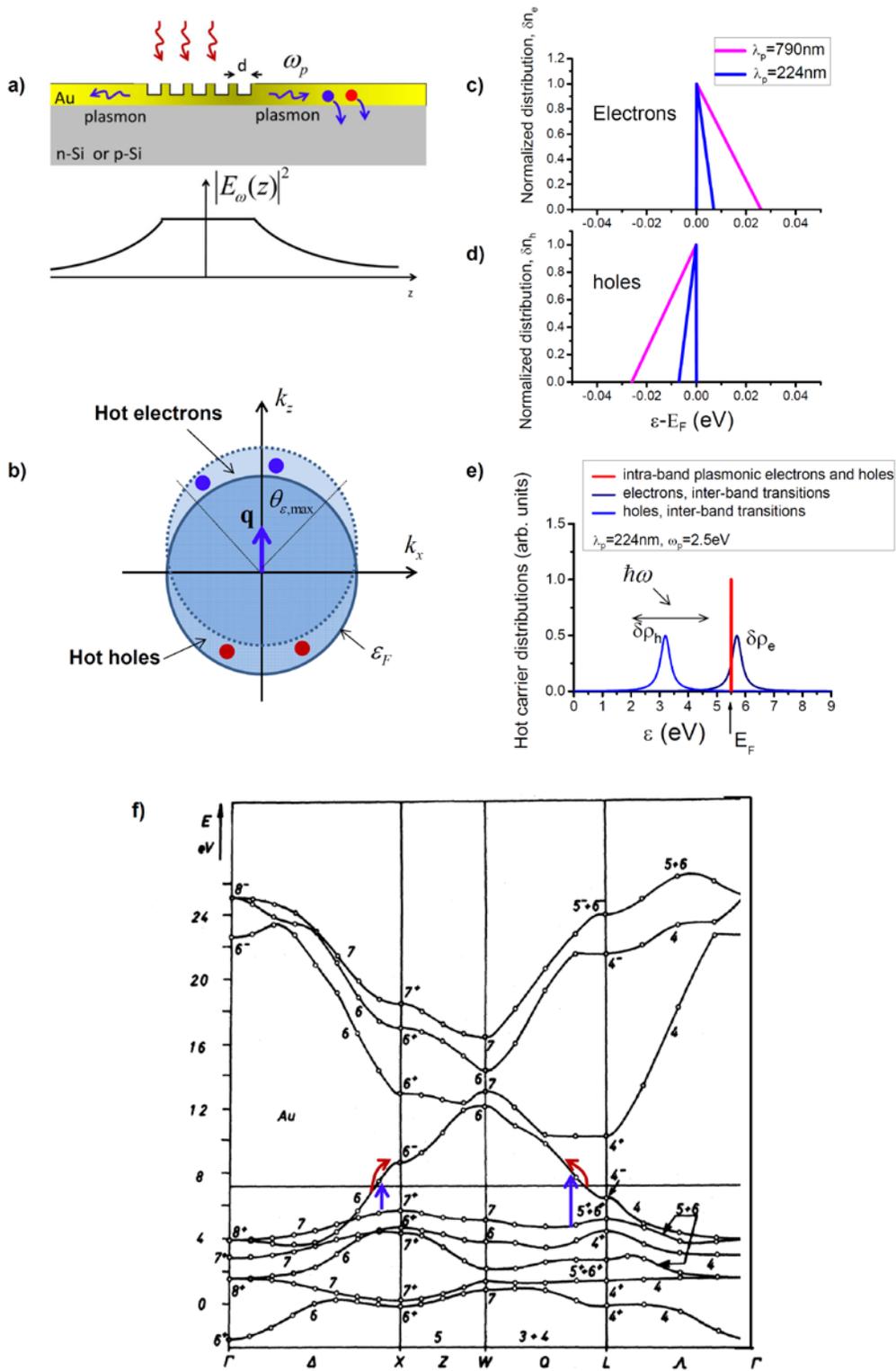

**Figure 2: a)** A planar plasmonic waveguide used to launch surface plasmon polaritons[41]. Plasmon can induce current to the substrate and be detected in this way[43]. **b)** Illustration of Fermi sea and hot carriers distributions in the k-space for the case of the intra-band transitions. To calculate the hot-electron distribution, we count the number of photo-excited electrons and holes in the indicated areas in the vicinity of the Fermi surface. **c)** and **d)** Distributions of electrons and holes for two plasmonic waves with wavelengths of 790nm and 224nm. Such waves can be excited in a plasmonic waveguide shown in (a). **e)** Schematics of the hot carrier distributions for the intra-band plasmonic carriers and for the inter-band electron and hole excitations. The energy interval of the distribution of intra-band plasmonic carriers is very narrow ($\sim 2\hbar q_p v_F$) and centered at the Fermi level. The inter-band transitions occupy wider energy intervals. **f)** Calculated band diagram of gold taken from Ref.40. Red arrows depict the intra-band plasmonic transitions, whereas the blue ones are the inter-band transitions. Adopted and reproduced with permission from Ref.40. Copyright 1971 by American Physical Society.

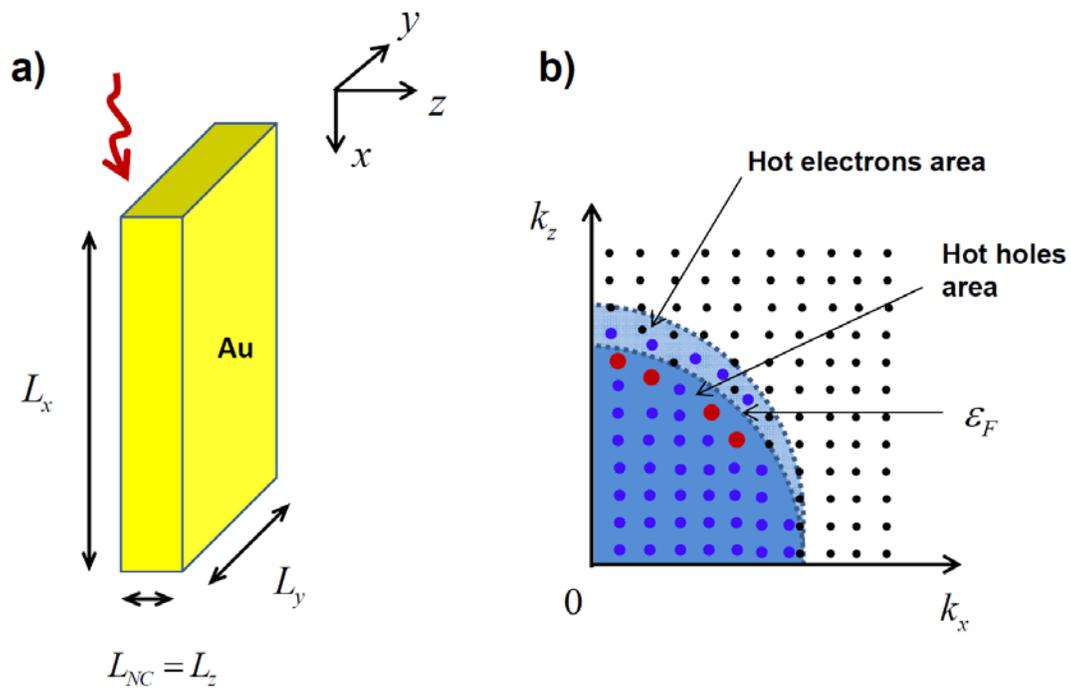

**Figure 3: a)** A model of plasmonic nano-antenna. **b)** The space of quantized states of electron in a NC. In the equilibrium, electrons occupy all states below the Fermi energy. Under excitation, the regions of hot electrons and holes become formed.

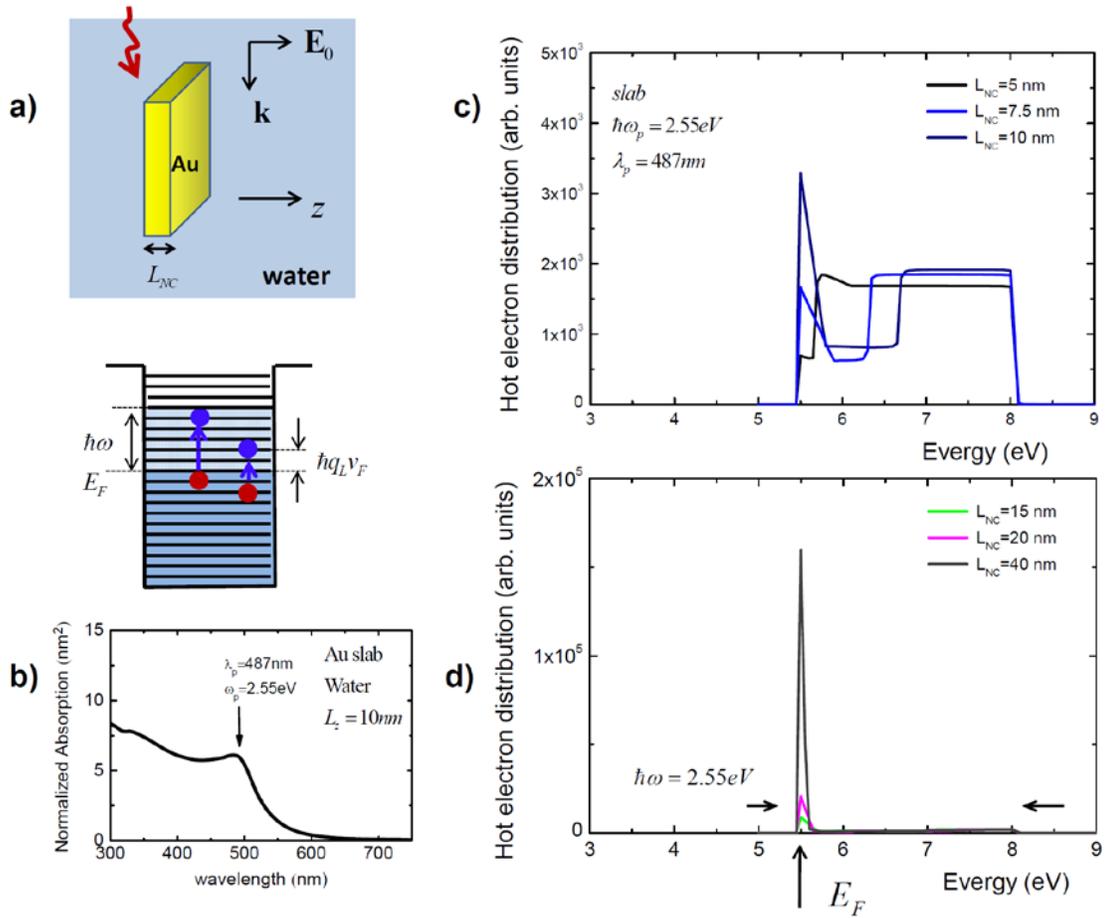

**Figure 4: a)** Model of platelet in water with $\mathbf{E}_0 \parallel \mathbf{z}$ and an energy diagram of electron states in a NC with important energy scales. **b)** Calculated absorption of an Au slab in water. **c)** and **d)** Calculated hot electron distributions for Au-platelets with various widths, $L_{NC} = L_z = 5, 7.5, 10, 15, 20, 30, 40$ nm. Amplitudes in all curves were not normalized and can be compared.

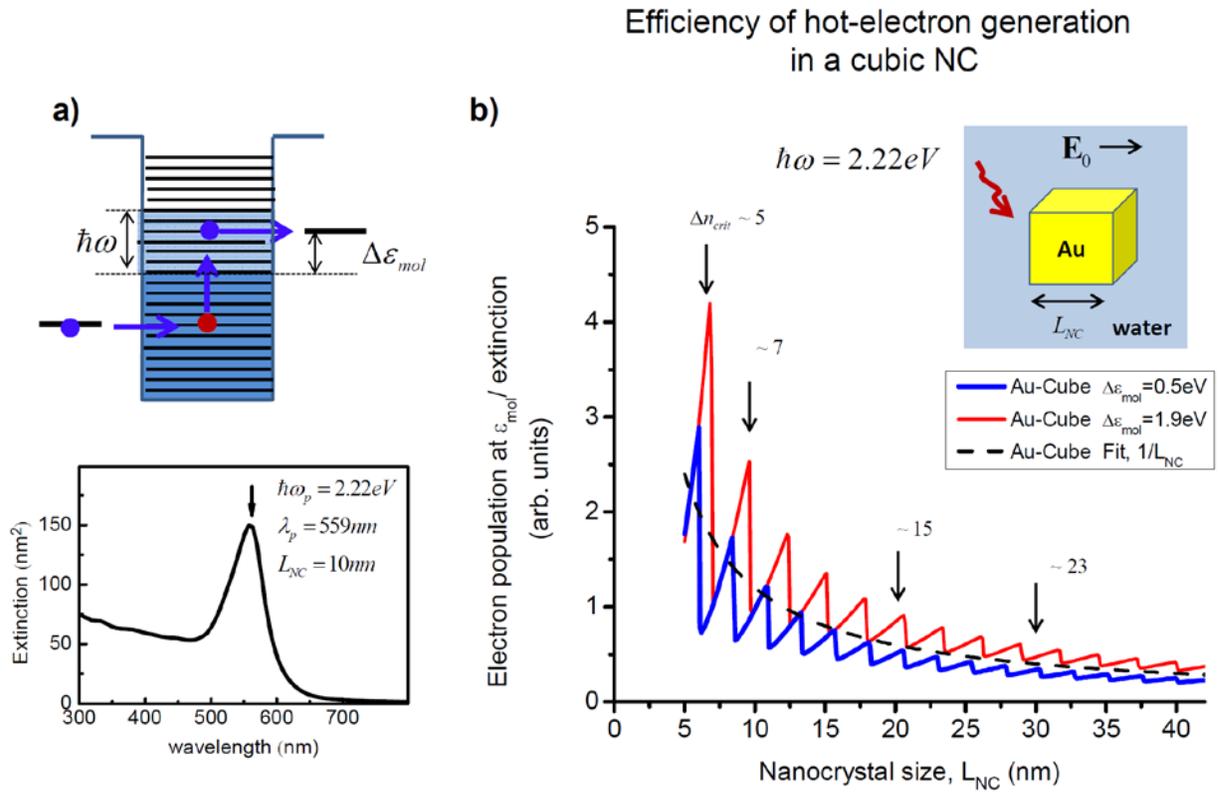

**Figure 5: a)** Schematics of the electron transfer processes between a NC and a molecule at the surface. Related energies are also shown. The absorption spectrum of Au cube calculated by the DDA method. **b)** Calculated efficiencies of hot-electron generation for molecules with $\Delta\varepsilon_{mol} = 0.5$ and $1.9\,eV$; the photon energy is in resonance with the dipolar plasmon mode. We also show the parameter $\Delta n_{crit} = \hbar\omega/\hbar q_L v_F$ which determines the efficiency of generation of high-energy carriers. Insert: Model of Au cube.

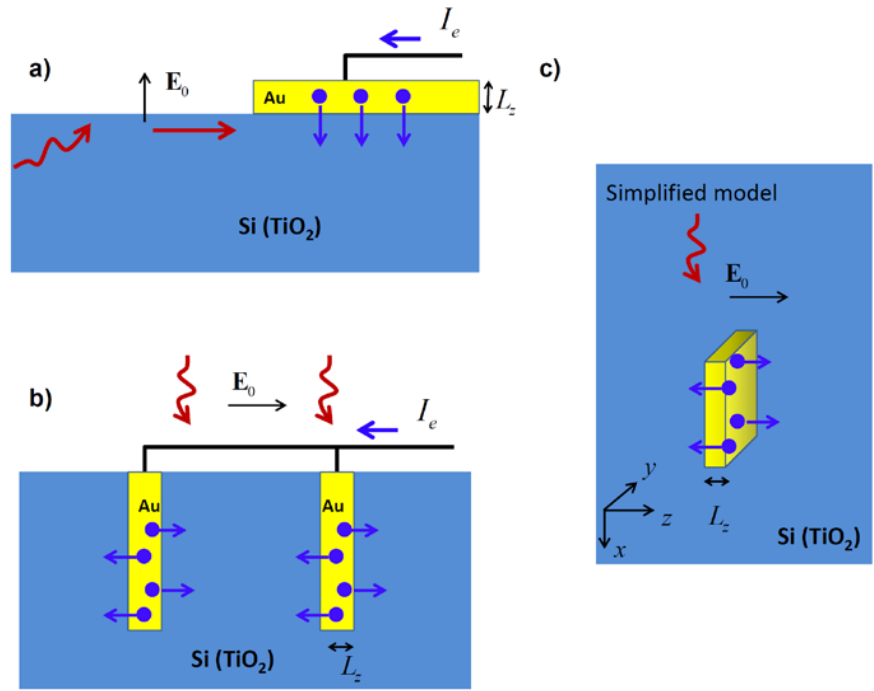

**Figure 6**: **a)-c)** Various geometries of plasmonic prism-like nano-antenna. In Figure (a), an evanescent wave in Si propagates along an interface and interacts with a metal platelet. In (b), metal nano-antennas are perpendicular to the surface and parallel to the direction of light with normal incidence. Figure (c) shows a simplified model used in the calculations.

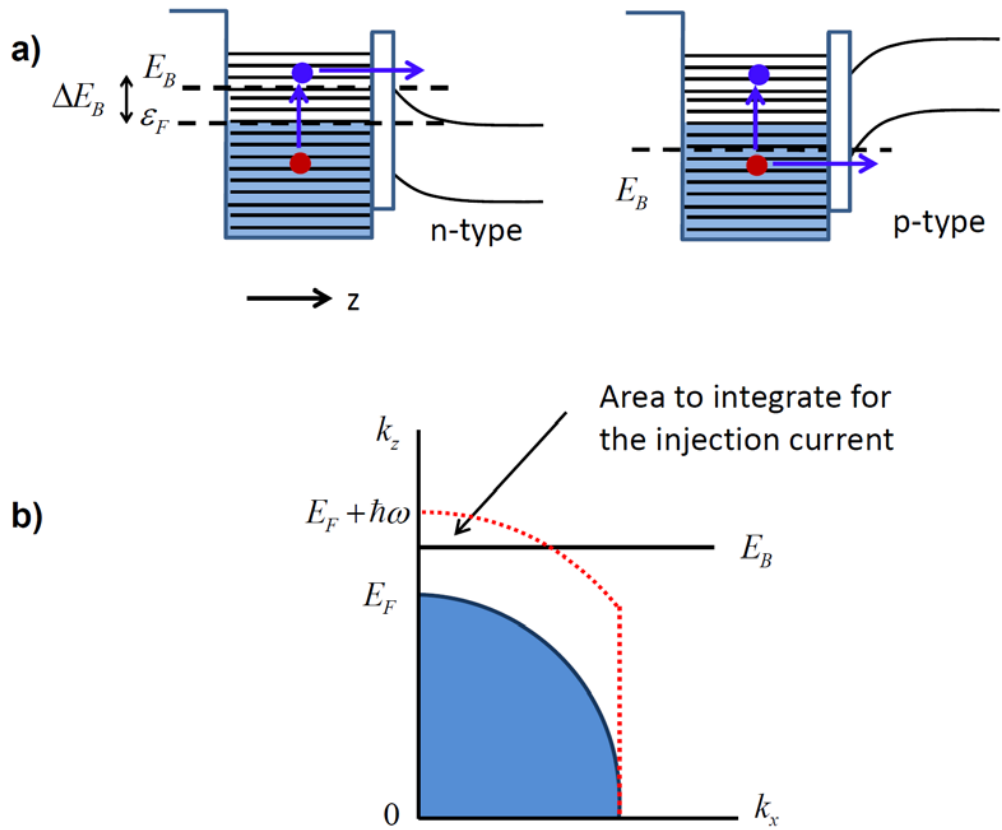

**Figure 7**: **a)** Energy diagrams showing injection of electrons and holes into a semiconductor through a Schottky barrier. **b)** Electron momentum space and the area of the space contributing to the injection process.

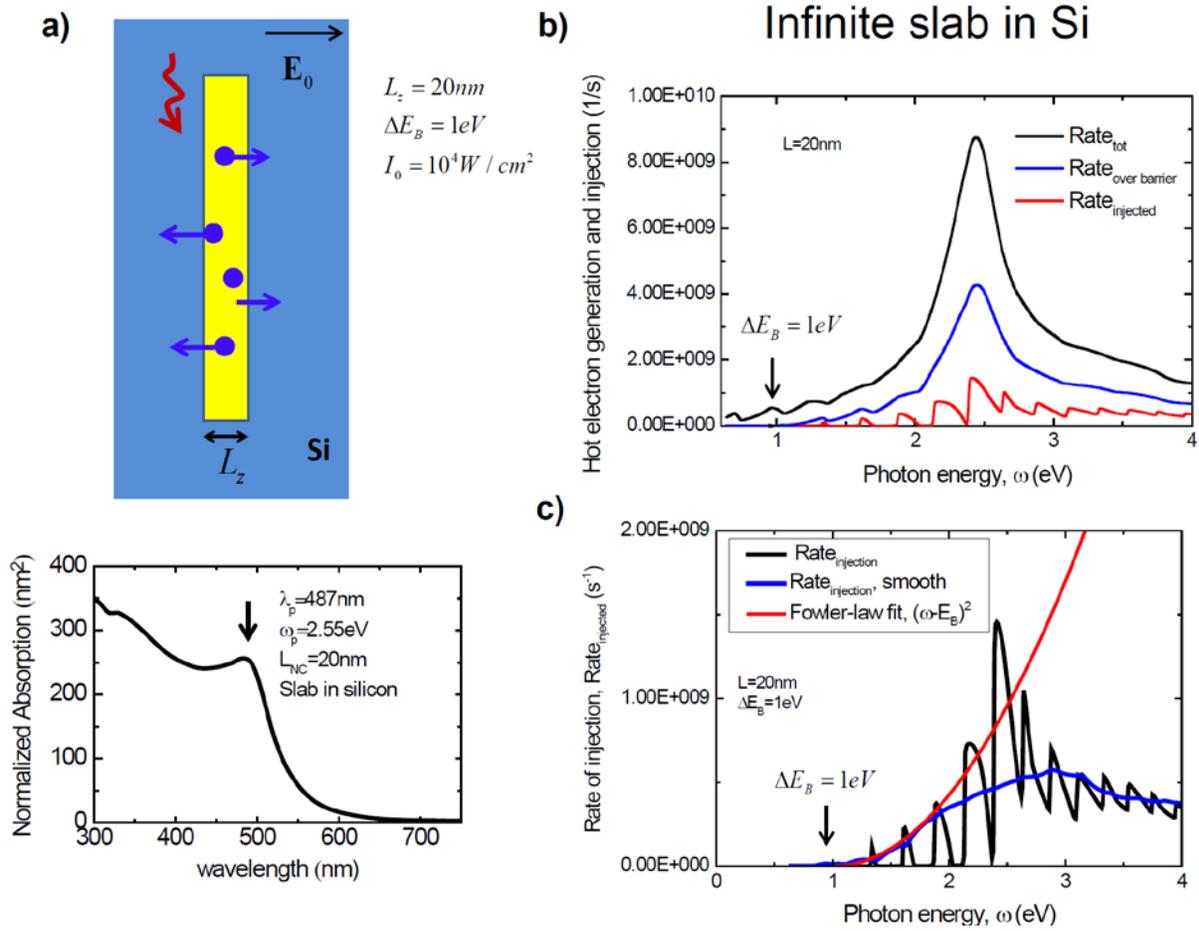

**Figure 8**: **a)** The model and absorption spectrum of Au slab. **b)** Rates of hot-electron generation and injection for a 20nm-slab. **c)** Rate of injection compared with the Fowler law. This graph also shows a smoothen curve.

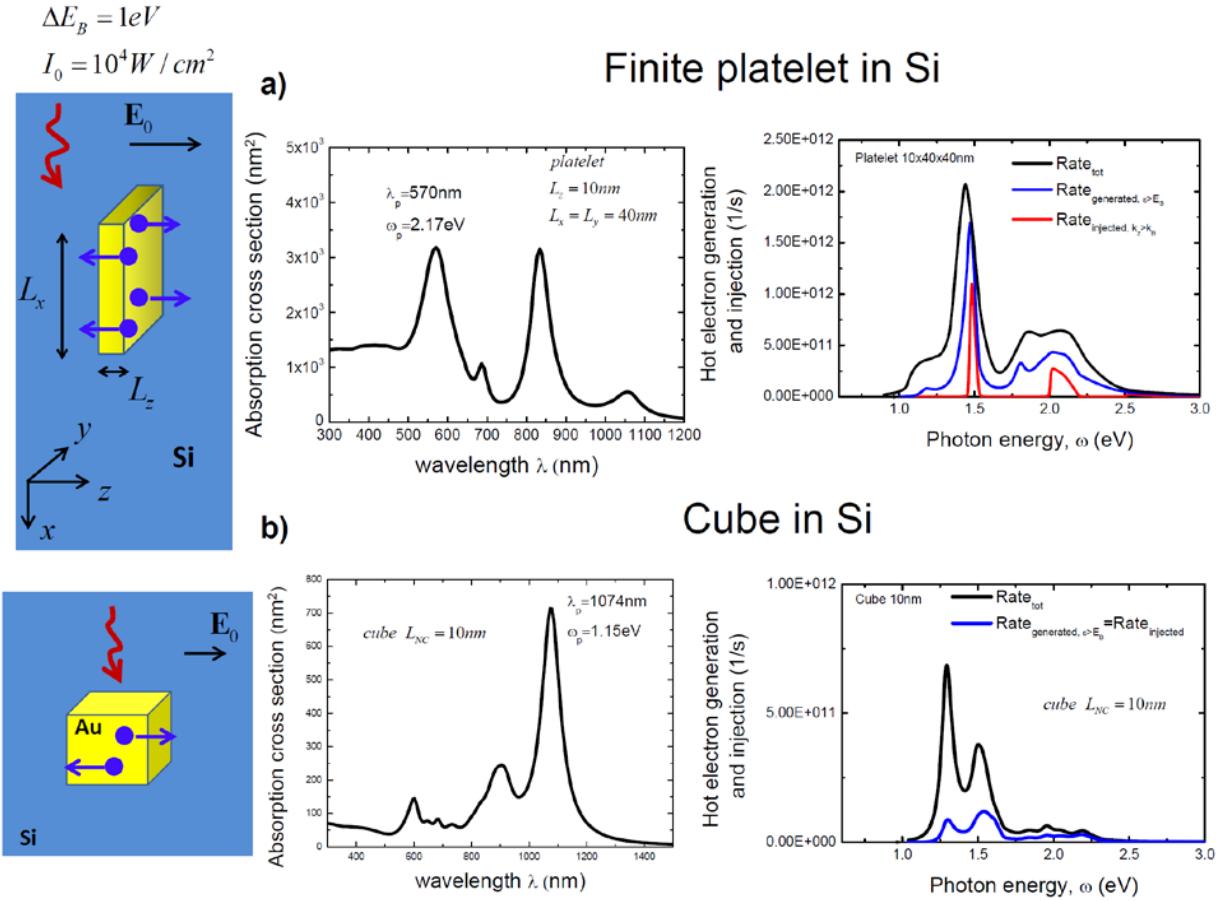

**Figure 9**: **a)** The absorption spectrum and the rates of generation of hot carriers in an Au platelet in Si. **b)** The same for a small Au cube in Si.

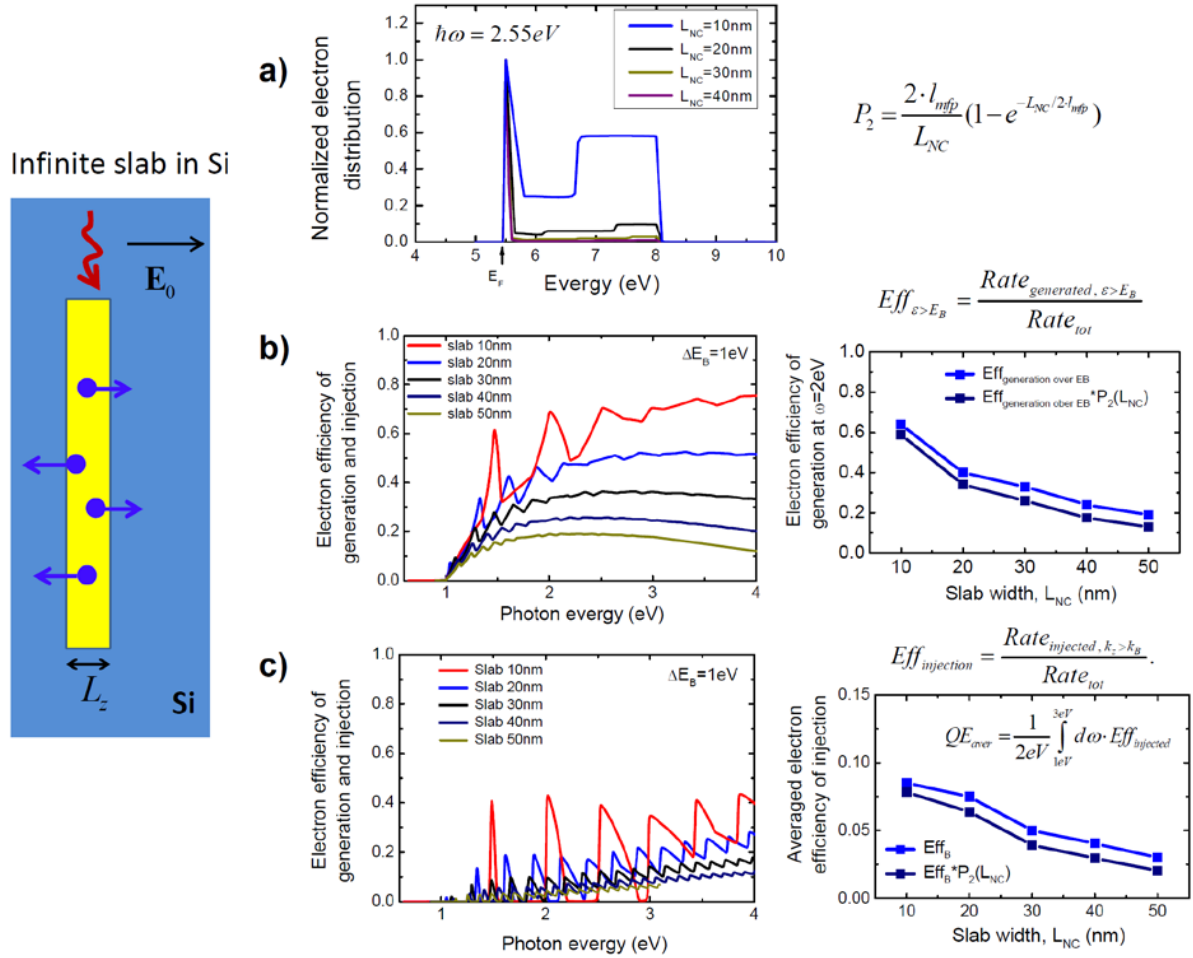

**Figure 10**: **a)**. Hot-electron distributions in Au slabs with various sizes. **b)** Electron efficiencies of generation of over-barrier carriers for slabs with various sizes. **c)** Electron efficiencies of injection.

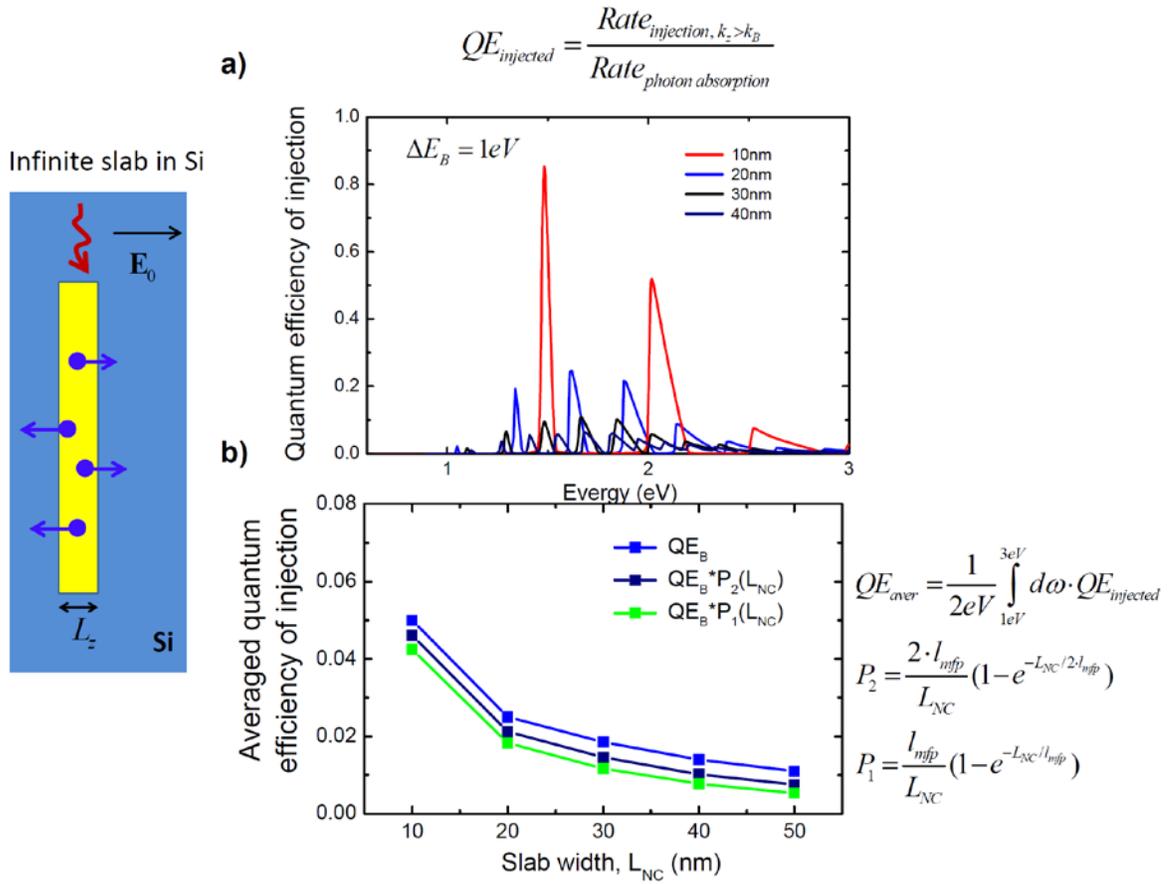

**Figure 11**: **a**). Quantum efficiencies of generation of over-barrier carriers for slabs with various sizes. **b**) Averaged quantum efficiencies of injection as functions of a width. These curves also include the effect of mean free path.

**TOC graphic:**

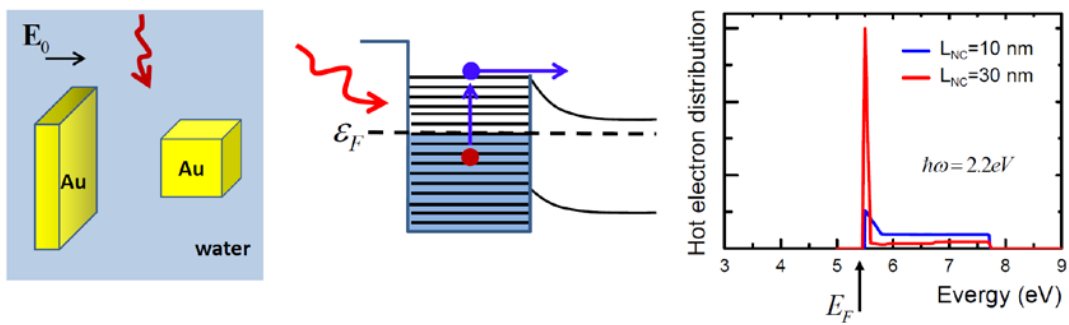